\documentclass[twocolumn,prl,superscriptaddress]{revtex4-2}
\usepackage[latin9]{inputenc}
\usepackage{verbatim}
\usepackage{amsmath}

\usepackage{amssymb}
\usepackage{graphicx}

\makeatletter
\usepackage{eqnarray}\usepackage{amsthm}
\usepackage{fancyhdr}
\usepackage{subfigure}\usepackage{epsfig}
\usepackage{epsfig}
\usepackage{bbm}
\usepackage{subfigure}
\usepackage{bm}
\usepackage{float}
\usepackage{xcolor}

\newcommand{\ep}{\boldsymbol{\epsilon}}
\newcommand{\DD}{\mathbf{D}}
\newcommand{\rr}{\boldsymbol{r}}
\newcommand{\nn}{\boldsymbol{n}}

\newcommand{\Gam}{\boldsymbol{\Gamma}}
\newcommand{\Lam}{\boldsymbol{\Lambda}}

\makeatother

\begin{document}
\title{Collisions enhance self-diffusion in odd-diffusive systems}

\author{Erik Kalz}
\affiliation{Leibniz-Institut f\"ur Polymerforschung Dresden, Institut Theorie der Polymere, 01069 Dresden, Deutschland}
\affiliation{Technische Universit\"at Dresden, Institut f\"ur Theoretische Physik, 01069 Dresden, Deutschland}

\author{Hidde Derk Vuijk}
\affiliation{Leibniz-Institut f\"ur Polymerforschung Dresden, Institut Theorie der Polymere, 01069 Dresden, Deutschland}

\author{Iman Abdoli}
\affiliation{Leibniz-Institut f\"ur Polymerforschung Dresden, Institut Theorie der Polymere, 01069 Dresden, Deutschland}

\author{Jens-Uwe Sommer}
\affiliation{Leibniz-Institut f\"ur Polymerforschung Dresden, Institut Theorie der Polymere, 01069 Dresden, Deutschland}
\affiliation{Technische Universit\"at Dresden, Institut f\"ur Theoretische Physik, 01069 Dresden, Deutschland}

\author{Hartmut L\"owen}
\affiliation{Heinrich Heine-Universit\"at  D\"usseldorf, Institut f\"ur Theoretische Physik II: Weiche Materie, 40225 D\"usseldorf, Deutschland}

\author{Abhinav Sharma}
\affiliation{Leibniz-Institut f\"ur Polymerforschung Dresden, Institut Theorie der Polymere, 01069 Dresden, Deutschland}
\affiliation{Technische Universit\"at Dresden, Institut f\"ur Theoretische Physik, 01069 Dresden, Deutschland}

\begin{abstract}
It is generally believed that collisions of particles reduce the self-diffusion coefficient. Here we show that in odd-diffusive systems, which are characterized by diffusion tensors with antisymmetric elements, collisions surprisingly can enhance the self-diffusion. In these systems, due to an inherent curving effect, the motion of particles is facilitated, instead of hindered by collisions leading to a mutual rolling effect. Using a geometric model, we analytically predict the enhancement of the self-diffusion coefficient with increasing density. This counterintuitive behaviour is demonstrated in the archetypal odd-diffusive system of Brownian particles under Lorentz force. We validate our findings by many body Brownian dynamics simulations in dilute systems.
\end{abstract}
\maketitle

Self-diffusion is related to the dynamics of a single particle, commonly referred to as the \emph{tracer particle}, in a homogeneous system of \emph{host particles}~\cite{ackerson1976correlations, ackerson1978correlations, dieterich1979memory, dhont1996introduction}. For purely repulsive interaction potentials, it is quite intuitive that the tracer particle is hindered in its motion by the host particles giving rise to a slowdown in the dynamics. The long-time self-diffusion coefficient $D_\mathrm{s}$ for a system of Brownian particles can be calculated exactly in the low-density limit as $D_\mathrm{s} = D_0 (1-\alpha\phi)$, where $\phi$ is the area fraction, $D_0$ is the diffusion coefficient at infinite dilution, and $\alpha$ is a numerical factor that depends on the nature of interactions ($\alpha = 2$ for hard-core interactions)~\cite{batchelor1976brownian,felderhof1978diffusion,jones1979diffusion,hanna1982self}. The reduction in the self-diffusion coefficient of the tracer particle with increasing density of host particles has been thoroughly demonstrated in experimental and computational studies~\cite{medina1988long, lowen1993long,imhof1995long,thorneywork2015effect}.

In this letter, we study self-diffusion coefficient in systems which are characterized by probability fluxes that are perpendicular to the density gradients. Analogous to odd-viscosity~\cite{avron1998odd,banerjee2017odd,reichhardt2021active}, recently such diffusive behaviour has been aptly termed as odd-diffusive \cite{hargus2021odd} and has attracted considerable attention \cite{chun2018emergence,vuijk2020lorentz,abdoli2020nondiffusive, vuijk2019anomalous, matsuyama2021anomalous,park2021thermodynamic, markovich2021odd, scheibner2020odd}.
We show that in odd-diffusive systems, collisions, instead of hindering the motion of the tracer particle, facilitate it resulting in an enhancement of the dynamics. Specifically, we demonstrate that in the low density limit, increasing the density of host particles leads to an increase in the self-diffusion coefficient of the tracer particle. Moreover, by tuning the odd-diffusivity, particles can be rendered dynamically invisible such that the tracer particle diffuses as a free particle.

\begin{figure}[t]
\includegraphics[clip, trim=4cm 2cm 4cm 2cm, width=\columnwidth]{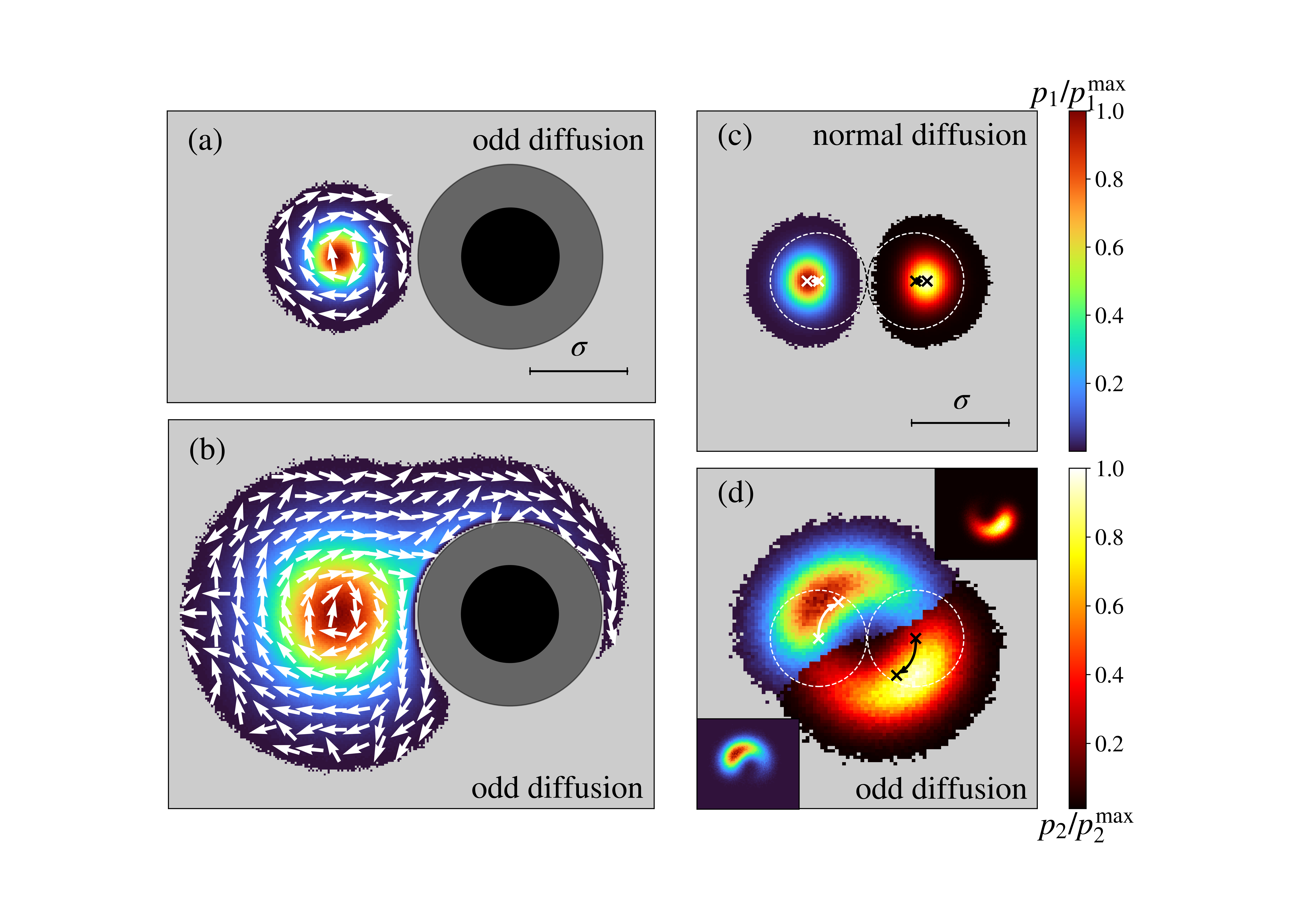}
\caption{Scaled density distribution ($p_1/p_1^\mathrm{max}$) of an odd-diffusive particle near a fixed particle of same diameter $\sigma$ at short (a) and long times (b). The curved probability fluxes (arrows) flow around the fixed particle in a preferred direction which can be flipped by reversing the odd-diffusivity parameter $\kappa$. Collision between two identical odd-diffusive particles therefore result in a probability flow around each other (d).  This "mutual rolling" of the particles around each other facilitates the motion resulting in an enhanced self-diffusion. This is in contrast to normal diffusive particles, where a collisions results in a reduced self-diffusion (c). The dashed circles represent the initial configuration and the crosses indicate the displacements of particles' centers. Insets in (d) show individual probability distributions of the two particles. The results are obtained from Brownian dynamics simulations of charged Brownian particles under Lorentz force with $\kappa = 5$.
}
\label{fig:density_flow}
\end{figure}

Odd-diffusive behavior emerges naturally in systems with broken time-reversal and parity symmetry. A charged Brownian particle in a magnetic field is a classical example of a system with broken time-reversal symmetry~\cite{chun2018emergence}. Other prominent examples of odd-diffusive systems are strongly damped particles subjected to Magnus~\cite{reichhardt2021active}, or Coriolis~\cite{kahlert2012magnetizing, hartmann2019self} forces or active chiral fluids~\cite{van2008dynamics, sevilla2016diffusion, kurzthaler2017intermediate, kummel2013circular}. In the limit of low persistence length, an active chiral particle follows curved trajectories, similar to the Brownian motion of a charged particle under a magnetic field.
 
The Fokker-Planck equation (FPE) for the probability density $P(\mathbf{r}, t)$ for an odd-diffusive particle reads as
\begin{equation}
     \frac{\partial P(\mathbf{r}, t)}{\partial t} = \nabla\cdot\left[\DD\ \nabla P(\mathbf{r}, t)\right], 
\end{equation}
where the diffusion tensor for two-dimensional isotropic systems can be written in the general form~\cite{chun2018emergence,hargus2021odd}
\begin{equation}
    \DD = D_0 \left(\mathbf{1} + \kappa \ep\right),
\label{diffusion_tensor}
\end{equation}
where $\mathbf{1} = \left(\begin{smallmatrix} 1 & 0 \\ 0 & 1 \end{smallmatrix}\right)$ is the identity tensor and $\ep = \left(\begin{smallmatrix} 0 & 1 \\ -1 & 0 \end{smallmatrix}\right)$ is the antisymmetric Levi-Civita symbol in two dimensions. $D_0$ is the normal diffusion coefficient of a particle which governs the usual diffusive fluxes parallel to density gradients. The odd-diffusive behaviour is characterized by the parameter $\kappa$ which governs fluxes in the direction perpendicular to the density gradients ~\cite{vuijk2019anomalous}. These fluxes have been called Lorentz fluxes in the context of a Brownian particle diffusing under the effect of Lorentz force. Since these fluxes are divergence free, they do not affect the density distribution of a single particle. However, in presence of boundaries, the Lorentz fluxes play an important role in the time evolution of density distributions~\cite{hargus2021odd}.

The surface of a fixed particle, for example, can be regarded as an impenetrable boundary for other particles. In Figs.~\ref{fig:density_flow}(a,b), we show the probability distribution and fluxes of an odd-diffusive particle near a fixed particle. At short times, the particle undergoes free diffusion, similar to a normal diffusive particle ($\kappa = 0$), see Fig.~\ref{fig:density_flow}(a). In contrast, the distribution of the particle at later times is strongly affected by the fixed particle, see Fig.~\ref{fig:density_flow}(b), where the effect of odd-diffusivity is strikingly evident. The curved probability fluxes flow around the fixed particle in a preferred direction. This phenomenon has no counterpart in a normal diffusive system. In fact, as we show below, the origin of the enhanced self-diffusion in odd-diffusive systems is fundamentally related to these curved fluxes.

We now consider two diffusing particles. From the perspective of the tracer particle, a host particle represents a moving boundary at which the flux of the tracer must vanish at all times. For particles performing normal diffusion ($\kappa = 0$), a collision is followed by the two particles moving away from each other (see Fig.~\ref{fig:density_flow}(c)). It is apparent that collisions hinder the diffusive exploration of space. Odd-diffusive particles, on the contrary, diffuse in a strikingly different way: when two particles collide, rather than blocking each other, they move around each other (see Fig.~\ref{fig:density_flow}(d)). In fact, for sufficiently large $\kappa$, a collision effectively facilitates the motion of particles around each other, an effect which we refer to as the "mutual rolling effect". 

We consider a model system of two distinguishable hard-core Brownian particles of diameter $\sigma$ in two dimensions. Ignoring hydrodynamic interactions, the FPE corresponding to the probability density of this system is
\begin{equation}
     \frac{\partial P(t)}{\partial t} = \nabla_1\cdot\left[\DD_1 \nabla_1 P(t)\right] + \nabla_2 \cdot\left[\DD_2\nabla_2 P(t)\right], 
\end{equation}
where $P(t) \equiv P(\rr_1,\rr_2,t)$. $\rr_1$ and $\rr_2$ are the position vectors of particle one and two, respectively. $\DD_1$ and $\DD_2$ are the corresponding odd-diffusion tensors (Eq.~\eqref{diffusion_tensor}). Since the particles cannot overlap, the FPE is defined only in the region $\Omega = \mathbb{R}^2 \times \mathbb{R}^2$\textbackslash$\mathcal{B}$ where $\mathcal{B} = \left\{(\rr_1,\rr_2) \in \mathbb{R}^2 \times \mathbb{R}^{2}; ||\rr_1 - \rr_2|| \leq \sigma\right\}$ is the forbidden area due to an overlap. The hard-core interactions impose a no-flux boundary condition on the moving boundary $\partial\mathcal{B}$
\begin{equation}
    \DD_1 \left[\nabla_1 P\right]\cdot\nn_1 + \DD_2 \left[\nabla_2 P\right]\cdot\nn_2 = 0,
    \label{boundarycondition}
\end{equation}
where $\nn_1$ and $\nn_2$ are outward unit normal vectors of the two particles such that $\nn_1 = -\nn_2$ on $\partial \mathcal{B}$. 

Our goal is to obtain an effective description for the marginal densities $p_1 \equiv \int_{\Omega(\rr_1)} \mathrm{d}\rr_2\ P(\rr_1,\rr_2,t)$ and $p_2 \equiv \int_{\Omega(\rr_2)}\mathrm{d}\rr_1\ P(\rr_1,\rr_2,t)$ for particles one and two, respectively. Here $\Omega(\mathbf{r}_i) = \mathbb{R}^2 \setminus \mathrm{B}_\sigma(\mathbf{r}_i)$, where $\mathrm{B}_\sigma(\mathbf{r}_i)$ is the disk of radius $\sigma$ centered at $\mathbf{r}_i, i \in \{1,2\}$. In the dilute regime, where only two-body collisions are relevant, we use an asymptotic method adapted from Bruna and Chapman~\cite{bruna2012excluded, bruna2012diffusion, bruna2015diffusion} which treats the effect of density perturbatively. A crucial step is the inclusion of the zero-flux boundary condition in Eq.~\eqref{boundarycondition}, which differs from the usual Neumann condition in normal diffusive systems. The detailed calculations are shown in the Supplementary Information (SI).

We generalise our description to include an arbitrary number of particles of each species in the low density limit. The effective FPEs for $p_i$ read as 
\begin{align}
     \frac{\partial p_i}{\partial t} &= \nabla\cdot \DD_i[\nabla p_i + (N_i - 1)\sigma^2\pi p_i\nabla p_i  \nonumber \\
     &\quad + N_j\pi \sigma^2 \left(\Lam_i p_i \nabla p_j - \Gam_i p_j \nabla p_i\right)],
\end{align}
where $N_i$ and $N_j$ denote the number of particles of species $i$ and $j$ with $(i,j) = (1,2)$ and $(2,1)$ and
\begin{align}
    \Lam_i &\equiv \mathbf{1} + \frac{\DD_1 + \DD_2}{\det(\DD_1 + \DD_2)}\DD_j,\\
    \Gam_i &\equiv \frac{\DD_1 + \DD_2}{\det(\DD_1 + \DD_2)}\DD_i.
\end{align}

\begin{figure}[t]
\centering
\includegraphics[clip, trim = 0cm 1.6cm 0cm 3cm, width=\columnwidth]{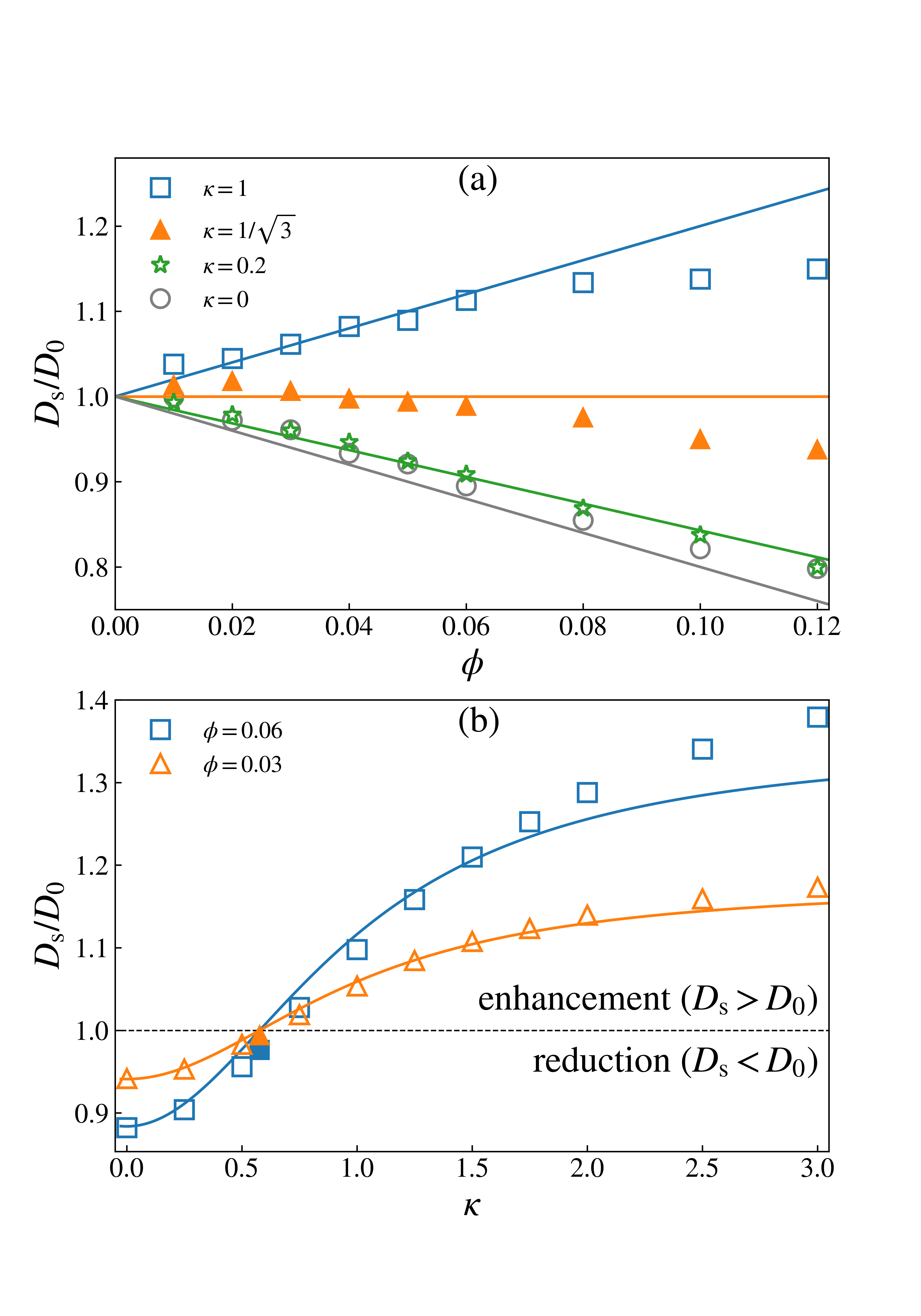}
\caption{(a) Reduced self-diffusion coefficient $D_\text{s}/D_0$ as a function of the area fraction $\phi$ for different values of the odd-diffusivity parameter $\kappa$. Symbols represent data from Brownian dynamics simulations.
Error bars are smaller than the symbols. Solid lines correspond to the analytical prediction of Eq.~\eqref{eq_self_diffusion}. For the critical value $\kappa_\mathrm{c} = 1/\sqrt{3}$ (filled symbols) particles are effectively invisible to each other. For $\kappa = 1\ (> \kappa_\mathrm{c})$ the self-diffusion coefficient increases with increasing $\phi$ and for $\kappa = 0.2\ (< \kappa_\mathrm{c})$ it decreases with increasing $\phi$. (b) $\kappa$-governed crossover from a reduced to an enhanced self-diffusion for two different area fractions. Self-diffusion coefficient for $\kappa = \kappa_\mathrm{c}$ is shown as filled symbols.
}
\label{fig:Ds}
\end{figure}
In our model, the two species can be differentiated via their $\kappa_i$ parameter and their diffusivity $D_0^{(i)}$. This makes the model well suited for studying diffusion of a tagged particle in presence of host particles as well as the collective diffusion of identical particles.

Let us consider that the two species have identical $\kappa$ and $D_0$, i.e., $\DD_1 = \DD_2 = \DD$. To obtain the self-diffusion coefficient, we set $N_1 = 1$ (tagged particle), $N_2 = N$ and define $\phi = N\pi\sigma^2p_2/4$ as the area fraction of host particles in two dimensions. The equation for the tagged particle reduces to
\begin{equation}
\label{eq_all_identical_particles}
\frac{\partial p_1(\rr,t)}{\partial t} = \nabla \cdot \DD\left[(\mathbf{1}-4\phi \Gam)\nabla p_1\right],
\end{equation}
where $\Gam = \Gam_1 = \Gam_2$ due to the identical diffusion matrices. Note that the probabilistic flux due to the odd part of the self-diffusion tensor $\DD(\mathbf{1}-4\phi\Gam)$ in Eq.~\eqref{eq_all_identical_particles} is divergence free and hence does not contribute to the mean-squared displacement of the particle. Hence, the self-diffusion coefficient is determined by the symmetric part alone and reads as 

\begin{equation}
\label{eq_self_diffusion}
D_\text{s} = D_0\left( 1 - 2 \phi \frac{1 - 3\kappa^2}{1 + \kappa^2}\right).
\end{equation}
Equation~\eqref{eq_self_diffusion} is the main result of this letter. It reduces to the well known expression for normal diffusive systems ($\kappa =0$) with hard-core interactions  $D_\mathrm{s} = D_0(1-2\phi)$~\cite{hanna1982self, felderhof1978diffusion, jones1979diffusion}. Our model generalizes this result. It predicts that for $\kappa < \kappa_{c} = 1/\sqrt{3}$, collisions with the host particles reduce the self-diffusion coefficient with respect to $D_0$. For $\kappa = \kappa_{c}$, the host particles become effectively invisible to the tagged particle, which diffuses with $D_\mathrm{s} = D_0$. The most interesting prediction of our model is that for $\kappa > \kappa_{c}$ the self-diffusion coefficient increases with increasing density of host particles (see Fig.~\ref{fig:Ds}(a)). In this regime, instead of hindering, collisions with the host particles facilitate the motion of the tagged particle (see Figs.~\ref{fig:density_flow}(c-f)) due to the mutual rolling effect. 
In Fig.~\ref{fig:Ds}(b), we show the variation of $D_\mathrm{s}$ with $\kappa$ for a fixed density of host particles. Increasing $\kappa$ results in a crossover from a reduced to an enhanced self-diffusion. Simulations reveal that odd-diffusivity enhances the self-diffusion coefficient relative to a normal diffusing system ($\kappa = 0$) at all densities (see SI). This suggests that the mutual rolling effect significantly compensates for the slowing down due to the many-body effects. In SI, we numerically show that for soft repulsive potentials our findings remain unaffected.

\begin{figure}[t]
\centering
\includegraphics[width=\columnwidth]{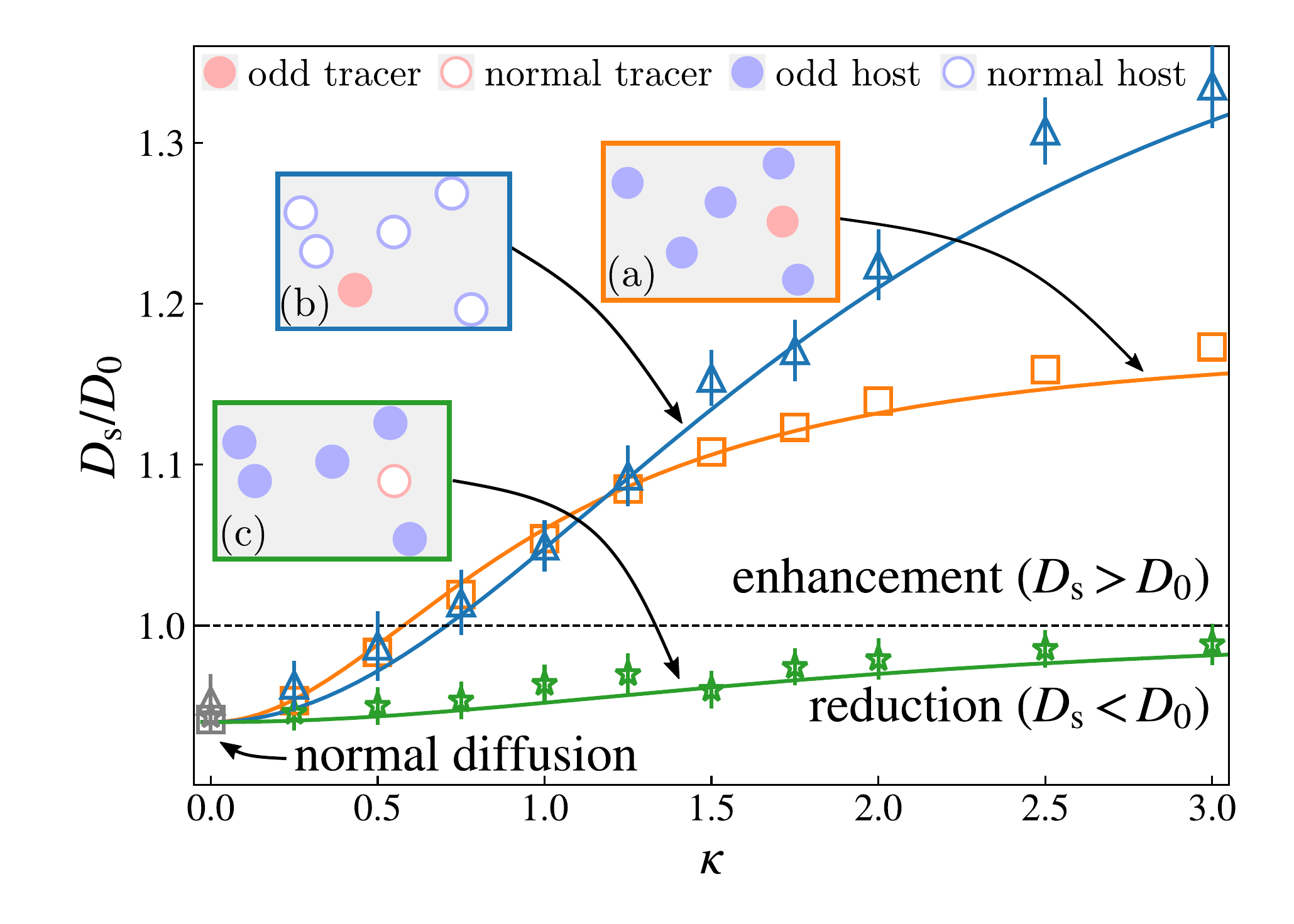}
\caption{Self-diffusion of tracer particle (red) in presence of host particles (blue) of area fraction $\phi = 0.03$. Symbols represent data from Brownian dynamics simulations.
(a) The tagged particle and the host particles are odd-diffusive. In this case, there is a crossover at a critical value of $\kappa_\mathrm{c} = 1/\sqrt{3}$ from reduction to enhancement of $D_\text{s}/D_0$. (b) The tagged particle is odd-diffusive and the host particles are normal. There is a crossover at $\kappa_\mathrm{c} = 1/\sqrt{2}$ again from reduction to enhancement. (c) For a normal tagged particle and odd-diffusive host particles, there is no enhancement of the self-diffusion. With increasing $\kappa$, the self-diffusion coefficient asymptotically approaches the ideal diffusivity, $D_\mathrm{s} = D_0$. The self-diffusion of normal diffusing particles ($\kappa = 0$) is shown in gray.}
\label{fig:self-diffusion_cases}
\end{figure}

We now consider the general case in which the tracer particle and the host particles are characterized by different values of the odd-diffusivity parameter $\kappa$.
Specifically, one can obtain the diffusion coefficient of a tracer particle diffusing in presence of host particles of different species. For $D_0$ the same for the tracer and the host particles, the self-diffusion coefficient of the tracer particle reads
\begin{equation}
\label{eq_general_self-diffusion}
D_\mathrm{s} = D_0 \left( 1 - 8\phi \frac{1 - \kappa_1 \kappa_2 - 2\kappa_1^2}{(\kappa_1 + \kappa_2)^2 + 4}\right),
\end{equation}
where $\kappa_1$ corresponds to the tagged particle and $\kappa_2$ to the host particles. Note that the result of the self-diffusion for identical particles (Eq.~\eqref{eq_self_diffusion}) is a special case corresponding to $\kappa_1 = \kappa_2$. Figure~\ref{fig:self-diffusion_cases}(a) depicts the scenario of identical odd-diffusive particles $(\kappa_1 = \kappa_2 = \kappa)$, for which the self-diffusion coefficient was shown in Fig.~\ref{fig:Ds}. Figure~\ref{fig:self-diffusion_cases}(b) shows $D_\mathrm{s}$ of an odd-diffusive particle ($\kappa_1 = \kappa$) in presence of normal diffusing host particles ($\kappa_2 = 0$). The self-diffusion coefficient $D_\mathrm{s} = D_0\left(1 - 8\phi\frac{1 - 2\kappa^2}{4 + \kappa^2}\right)$ shows a crossover from reduction to enhancement at $\kappa_\mathrm{c} = 1/ \sqrt{2}$ due to collisions.
For an odd-diffusive tracer particle for both, (a) and (b), a collision with the host particle facilitates its motion giving rise to an enhancement of the diffusion coefficient. Moreover, that (b) overtakes (a) for $\kappa > \sqrt{7/5}$ raises the question whether in this regime the diffusion of an odd particle is most efficient within (normal) obstacles. In contrast to the enhancement, Fig.~\ref{fig:self-diffusion_cases}(c) shows the case of an normal diffusing tracer particle ($\kappa_1 = 0$) in presence of odd-diffusive host particles ($\kappa_2 = \kappa$) for which $D_\mathrm{s} = D_0\left(1 - \frac{8\phi}{4 + \kappa^2}\right)$ does not show an enhancement of the self-diffusion coefficient. In this case, $D_\mathrm{s}$ approaches $D_0$ asymptotically. Nevertheless, odd-diffusivity of the host particles gives rise to a faster diffusion of the tracer in comparison to normal host particles.
The theoretical predictions are validated by Brownian dynamics simulations of hard-core interacting particles diffusing under the effect of the Lorentz force~\cite{Note1}. 

The different scenarios above highlight that only odd-diffusive tracer particles benefit from collisions with the host particles, in agreement with the physical mechanism suggested in Fig.~\ref{fig:density_flow}. The mutual rolling effect gives rise to unusual force autocorrelation as we show in SI in the low density limit. While the autocorrelation is a positive monotonically decaying function of time for normal diffusing ($\kappa =0$) particles~\cite{hanna1981velocity}, it exhibits zero crossing in time for odd-diffusive particles.

A collision helps the tracer particle escape local caging by the host particles resulting in enhanced exploration of space. From the perspective of the whole species, however, nothing has changed, only two particles have interchanged positions due to a collision. This suggests that the collective diffusion coefficient is unaffected by odd-diffusivity. By untagging the tracer particle in Eq.~\eqref{eq_all_identical_particles}, the inter-species terms drop out and we obtain a single-species system, with $N+1$ identical particles characterized by the collective diffusion coefficient

\begin{equation}
\label{eq_collective_diffusion}
D_\text{c} = D_0\left( 1 + 4 \phi\right), 
\end{equation}
which is indeed independent of $\kappa$ and has the same expression as in normal diffusive systems~\cite{berne2000dynamic, bruna2012diffusion}. We note that $D_\text{s} > D_\text{c}$ for large $\kappa$ which might have interesting implications for density relaxation in odd-diffusive fluids.

Self-diffusion is affected by both direct and hydrodynamic interactions~\cite{santana2005short,michailidou2009dynamics}. Whereas hydrodynamic interactions can enhance the self-diffusion coefficient~\cite{zahn1997hydrodynamic,mcphie2007long}, direct interactions always reduce the self-diffusion in ordinarily diffusing systems. In this study, we ignored the hydrodynamic interactions and showed that in odd-diffusive systems, interactions can enhance the self-diffusion coefficient. 

Mesoscopic overdamped systems of colloidal spinners (magnetic dipoles), when exposed to a viscoeleastic solvent exhibit a Magnus force which can be tuned by the spinning frequency~\cite{xin}. With $\kappa$ of order unity and above possible, this could be a promising realization for odd-diffusive systems. Another possible realization is millimeter-sized granules, which beget high charges when exposed to a vibrating substrate due to triboelectric effects~\cite{kaufman2009phase_soft,kaufman2009phase}.
For a  millimeter sized sphere where inertia can almost be ignored, with the surface charge density $\sigma = 1$ nm$^{-2}$, the viscosity $\eta \approx 10^{-4}$ Pa s (Propylene at room temperature) and $B = 1$ T, one obtains $\kappa \approx 1$. A plausible experimental setup can also be realised in dusty plasma which can be almost overdamped for the high density of the ambient gas. Large magnetic fields exceeding $10^4$ T can be effectively realised using non-inertial rotating frames~\cite{kahlert2012magnetizing, hartmann2019self} at which $\kappa$ of order one is in reach for millimeter-sized dust particles.
Furthermore the mutual rolling effect should be verifiable in self-spinning granules~\cite{scholz2017velocity, cafiero2002rotationally, nguyen2014emergent}, in chiral colloidal microswimmers~\cite{van2008dynamics, sevilla2016diffusion, kurzthaler2017intermediate, kummel2013circular, kokot2017active}, even in rotating molecular motors ~\cite{sumino2012large, tabe2003coherent} and vortex fluids~\cite{petroff2015fast, riedel2005self}.

The enhancement of self-diffusion is reminiscent of Taylor dispersion~\cite{aris1956dispersion} in which flow along a direction affects diffusion along the orthogonal direction. This is phenomenologically similar to odd-diffusivity induced mutual rolling of two colliding particles. It would be interesting to investigate Taylor dispersion in a dilute suspension of odd-diffusive particles. Our findings might also be applicable to systems with Magnus forces where dragging a probe particle was found to speed up with increasing system density~\cite{reichhardt2021active, brown2018effect}. Recently wiggling nanopores have been shown to enhance the diffusion coefficient of a particle ~\cite{marbach2018transport}. Surprisingly, the enhancement has exactly the same functional form as in Eq.\eqref{eq_self_diffusion}. Additional work is needed to investigate whether the similarity extends beyond the mathematical formalism employed in our work and fluctuating nanopores.
 
Finally, in general, the tunability of odd-diffusivity by external fields offers intriguing options to pave the way towards novel functional metamaterials with fascinating unusual dynamics. For example, particles can be rendered "dynamically invisible" though they are actually colliding. 

A.S. and H.L. acknowledge support by the Deutsche Forschungsgemeinschaft (DFG) within the projects SH 1275/3-1 and LO 418/25-1. J.-U. S. thanks the cluster of excellence "Physics of Life" at TU Dresden for its support.

\bibliographystyle{unsrt}

\end{document}


\title{Supplementary Information: Collisions enhance self-diffusion in odd-diffusive systems}

\author{Erik Kalz}
\affiliation{Leibniz-Institut f\"ur Polymerforschung Dresden, Institut Theorie der Polymere, 01069 Dresden, Deutschland}
\affiliation{Technische Universit\"at Dresden, Institut f\"ur Theoretische Physik, 01069 Dresden, Deutschland}

\author{Hidde Derk Vuijk}
\affiliation{Leibniz-Institut f\"ur Polymerforschung Dresden, Institut Theorie der Polymere, 01069 Dresden, Deutschland}

\author{Iman Abdoli}
\affiliation{Leibniz-Institut f\"ur Polymerforschung Dresden, Institut Theorie der Polymere, 01069 Dresden, Deutschland}

\author{Jens-Uwe Sommer}
\affiliation{Leibniz-Institut f\"ur Polymerforschung Dresden, Institut Theorie der Polymere, 01069 Dresden, Deutschland}
\affiliation{Technische Universit\"at Dresden, Institut f\"ur Theoretische Physik, 01069 Dresden, Deutschland}

\author{Hartmut L\"owen}
\affiliation{Heinrich Heine-Universit\"at  D\"usseldorf, Institut f\"ur Theoretische Physik II: Weiche Materie, 40225 D\"usseldorf, Deutschland}

\author{Abhinav Sharma}
\affiliation{Leibniz-Institut f\"ur Polymerforschung Dresden, Institut Theorie der Polymere, 01069 Dresden, Deutschland}
\affiliation{Technische Universit\"at Dresden, Institut f\"ur Theoretische Physik, 01069 Dresden, Deutschland}

\maketitle
In this Supplementary Information we present a method to obtain an effective time-evolution equation on the one-body level for hard-core interacting spherical Brownian particles in an odd-diffusive system, where we start the analysis with the Fokker-Planck equation for the joint distribution. Rather than dealing with the interactions via their (singular) interaction potential, we adopt the method developed by Bruna and Chapman \cite{bruna2012excluded, bruna2012diffusion}, who included the interactions via impenetrable effective moving boundaries (Section \ref{section_model}).  Additionally we give an outline to an alternative theory to deal with odd-diffusive systems, which in comparison to the main theory is restricted to equilibrium systems (Section \ref{section_outline}). We explain the details and characteristics of the simulations we performed (Section \ref{section_simulations}). And We give additional numerical data for soft interacting particles (Section \ref{section_soft_interactions}) and high densities of a hard-sphere system (Section \ref{section_high_densities}).

\section{Model}
\label{section_model}

\subsection{Time Evolution Equation}

\subsubsection{Equation for joint probability}
The time-evolution equation for the joint probability distribution $P(t) \equiv P(\mathbf{r}_1, \mathbf{r}_2, t)$ for two particles of diameter $\sigma$ located at $\mathbf{r}_1$ and $\mathbf{r}_2$ undergoing overdamped diffusion in an odd-diffusive system reads

\begin{equation}
\label{general_FP_eq}
\frac{\partial P(t)}{\partial t} = \nabla_1 \cdot \left[\mathbf{D}_1\ \nabla_1 P(t) \right] + \nabla_2 \cdot \left[\mathbf{D}_2\ \nabla_2 P(t) \right].
\end{equation}
Here $\nabla_i$ represents the partial derivative with respect to the position of particle with label $i$, $i\in \{1,2\}$, and $\mathbf{a} \cdot \mathbf{b}$ denotes the inner product of two vectors $\mathbf{a}, \mathbf{b}$. The following convention is followed throughout: in Cartesian coordinates the $x$-component of the vector $\mathbf{D}\ \nabla P$ reads as $(\mathbf{D}\ \nabla P)_x = \mathbf{D}_{xx} \partial_x P + \mathbf{D}_{xy} \partial_y P$. The divergence of the vector field is given as $\nabla \cdot \mathbf{D}\ \nabla P = \partial_x (\mathbf{D}\ \nabla P)_x + \partial_y (\mathbf{D}\ \nabla P)_y$.

The diffusion of the two particles is characterized by the general diffusion tensor for isotropic systems in two dimensions, which reads
\begin{equation}
\label{diffusion_tensor}
\mathbf{D}_i = D_0^{(i)} \left(\mathbf{1} + \kappa_i  \boldsymbol{\epsilon}\right),
\end{equation} 
with $\mathbf{1} = \left(\begin{smallmatrix}1 & 0\\ 0 & 1 \end{smallmatrix}\right)$ as the identity tensor and $\boldsymbol{\epsilon} = \left(\begin{smallmatrix}0 & 1\\ -1 & 0 \end{smallmatrix}\right)$ as the anti-symmetric Levi-Civita symbol in two dimensions. Here $D_0^{(i)}$ is the bare diffusivity of particle $i$, which governs the ordinary diffusive flux parallel to density gradients. The odd-diffusion behaviour of particle $i$ is characterized by the parameter $\kappa_i$, giving rise to probability fluxes perpendicular to density gradients. Consequently, $\kappa = 0$ corresponds to an ordinarily diffusing particle. The diffusion tensor here reduces to $\mathbf{D} = D_0^{(j)}\ \mathbf{1}$. 

\begin{figure}[t]
\centering
\includegraphics[width=0.5\columnwidth]{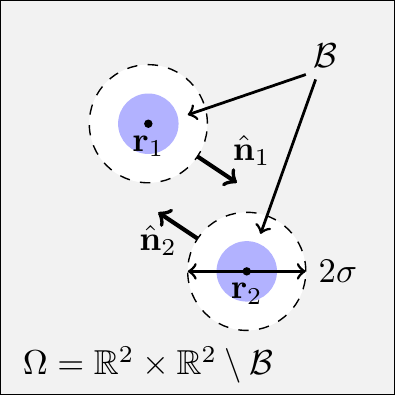}
\caption{The excluded volume $\mathcal{B}$ (white area), which two hard-core interacting spherical Brownian particles (blue) of diameter $\sigma$ carry with respect to the others centre-coordinate. The equation governing the time-evolution of the probability distribution of the two particles only is defined in the gray area $\Omega$, where there is a no-flux boundary condition defined on the dashed lines, the collision surface. $\hat{\mathbf{n}}_1$ and $\hat{\mathbf{n}}_2$ are the outward unit normal vectors of particles one and two, respectively. They live on $\partial\Omega$.}
\label{fig:sketch_ev}
\end{figure}

The particles interact with each other via a hard-core interaction. Following Refs.~\cite{bruna2012excluded, bruna2012diffusion}, the interaction is not build in the time-evolution in Eq.~\eqref{general_FP_eq} due to an interaction potential, but via the space in which Eq.~\eqref{general_FP_eq} is defined.
The finite area of one particle represents an excluded area for the others particles center coordinate. Hence Eq.~\eqref{general_FP_eq} is not defined in whole $\mathbb{R}^2 \times \mathbb{R}^2$, but instead in a reduced form $\Omega \equiv \mathbb{R}^2 \times \mathbb{R}^2 \setminus \mathcal{B}$. Here $\mathcal{B} = \{ (\mathbf{r}_1, \mathbf{r}_2) \in \mathbb{R}^2 \times \mathbb{R}^2; ||\mathbf{r}_1 - \mathbf{r}_2|| \leq \sigma\}$ is the set of all forbidden configurations corresponding to an overlap of the particles. This excluded area is illustrated in Fig.~\ref{fig:sketch_ev}. 

Defining an excluded area for the other particles center coordinate, in contrast to common approaches, we treat the hard-core interactions between the particles in a geometrical sense. The excluded-area interactions define a no-flux boundary condition on the so called collision surface $\partial\Omega = \partial \mathcal{B}$ for the time-evolution equation of the joint probability density in Eq.~\eqref{general_FP_eq}

\begin{equation}
\label{bc_for_general_FP_eq}
\left[\mathbf{D}_1\ \nabla_1 P(t) \right] \cdot \hat{\mathbf{n}}_1 + \left[\mathbf{D}_2\ \nabla_2 P(t) \right] \cdot \hat{\mathbf{n}}_2 = 0 \quad \text{on}\ \partial\Omega,
\end{equation}
where $\hat{\mathbf{n}}_i$ is the normalized component of the outward unit normal vector corresponding to the $i$th particle, i.e. $\hat{\mathbf{n}} = (\mathbf{n}_1, \mathbf{n}_2)$, see Fig.~\ref{fig:sketch_ev}. Note here that $\hat{\mathbf{n}}_1 = - \hat{\mathbf{n}}_2$ on $\partial \mathcal{B}$. 

We are interested in an effective time-evolution equation on the one-body level for the densities

\begin{equation}
p_1(\mathbf{r}_1,t) = \int_{\Omega(\mathbf{r}_1)} \mathrm{d}\mathbf{r}_2\ P(\mathbf{r}_1, \mathbf{r}_2, t)
\end{equation}
and 

\begin{equation}
p_2(\mathbf{r}_2,t) = \int_{\Omega(\mathbf{r}_2)} \mathrm{d}\mathbf{r}_1\ P(\mathbf{r}_1, \mathbf{r}_2, t)
\end{equation}
as the one-body densities for particle one and two, respectively.  Here $\Omega(\mathbf{r}_i) = \mathbb{R}^2 \setminus \mathrm{B}_\sigma(\mathbf{r}_i)$, where $\mathrm{B}_\sigma(\mathbf{r}_i)$ is the disk of radius $\sigma$ around $\mathbf{r}_i$, $i \in\{1,2\}$. This reduced configuration space respects that we integrate out one particle, when effectively fixing the other. In this integration we need to take the boundary condition of Eq.~\eqref{bc_for_general_FP_eq} into account when integrating Eq.~\eqref{general_FP_eq} over the others particles coordinate.
For illustration purposes we demonstrate the derivation for the time-evolution equation for the one-body density distribution of particle one, $p_1$. With an interchange of all particle labels, the time-evolution equation for the other particle can be easily obtained.

\subsubsection{Equation for one-body probability}
We want to integrate out the effect of the second particle on the first to obtain a time-evolution equation for $p_1$. Therefore we effectively fix particle one and as before denote the reduced configuration space for the coordinate of particle two as $\Omega(\mathbf{r}_1)$. Integrating the left-hand side of Eq.~\eqref{general_FP_eq} over $\Omega(\mathbf{r}_1)$ results in

\begin{equation}
\label{time_derivative}
\int_{\Omega(\mathbf{r}_1)} \mathrm{d}\mathbf{r}_2\ \frac{\partial}{\partial t}P(\mathbf{r}_1, \mathbf{r}_2, t) = \frac{\partial p_1(\mathbf{r}_1 t)}{\partial t}.
\end{equation}
When integrating the right-hand side of Eq.~\eqref{general_FP_eq} over the reduced configuration space $\Omega(\mathbf{r}_1)$, we find two distinct contributions. For one, we can apply the divergence theorem, since both the differentiation and the integration, are with respect to the coordinates of the second particle.

\begin{multline}
\int_{\Omega(\mathbf{r}_1)} \mathrm{d}\mathbf{r}_2\ \nabla_2 \cdot \left[\mathbf{D}_2\ \nabla_2 P(t) \right] = \\ = \int_{\partial\mathrm{B}_\sigma(\mathbf{r}_1)} \mathrm{d}\mathcal{S}_2\ \hat{\mathbf{n}}_2 \cdot \left[\mathbf{D}_2\ \nabla_2 P(t) \right],
\end{multline}
where $\mathrm{d}\mathcal{S}_2\ \hat{\mathbf{n}}_2$ is the directed surface element for the excluded area of particle two and we used that $\partial \Omega(\mathbf{r}_1) = \partial\mathrm{B}_\sigma(\mathbf{r}_1)$. On this result we can use the no-flux boundary condition of Eq.~\eqref{bc_for_general_FP_eq} to obtain

\begin{align}
\label{worked_in_bc}
\int_{\partial\mathrm{B}_\sigma(\mathbf{r}_1)} \mathrm{d}\mathcal{S}_2\ \hat{\mathbf{n}}_2 \cdot & \left[\mathbf{D}_2\ \nabla_2 P(t) \right] = \nonumber \\
& = - \int_{\partial\mathrm{B}_\sigma(\mathbf{r}_1)} \mathrm{d}\mathcal{S}_2\ \hat{\mathbf{n}}_1 \cdot \left[\mathbf{D}_1\ \nabla_1 P(t) \right] \nonumber\\
& = \int_{\partial\mathrm{B}_\sigma(\mathbf{r}_1)} \mathrm{d}\mathcal{S}_2\ \hat{\mathbf{n}}_2 \cdot \left[\mathbf{D}_1\ \nabla_1 P(t) \right],
\end{align}
where we used that $\hat{\mathbf{n}}_2 = -\hat{\mathbf{n}}_1$ on $\partial\mathrm{B}_\sigma(\mathbf{r}_1)$ for the last line.

Because for the other contribution the integration and the partial differentiation are not with respect to the same particle, the integral cannot be evaluated with the use of the divergence theorem, but is instead given as

\begin{multline}
\label{applied_generalized_RTT}
\int_{\Omega(\mathbf{r}_1)} \mathrm{d}\mathbf{r}_2\ \nabla_1 \cdot \left[\mathbf{D}_1\ \nabla_1 P(t) \right] = \nabla_1 \cdot \left[ \mathbf{D}_1 \nabla_1\ p_1(\mathbf{r}_1, t) \right] \\
- \int_{\partial\mathrm{B}_\sigma(\mathbf{r}_1)} \mathrm{d}\mathcal{S}_2\ \hat{\mathbf{n}}_2 \cdot \left[ \left(\mathbf{D}_1 + \mathbf{D}_1^\text{T}\right) \nabla_1 P(t) + \mathbf{D}_1^\text{T}\ \nabla_2 P(t) \right],
\end{multline}
where the superscript $(\cdot)^\text{T}$ denotes the transpose of a matrix. For a justification of Eq.\eqref{applied_generalized_RTT} see the Section \ref{section_justification_of_Eq_9}.

Collecting results, we obtain an effective time-evolution equation for the density $p_1$ of the first particle with the collision effects build in an integral

\begin{multline}
\label{p1_equation_with_collision_integral}
\frac{\partial p_1(\mathbf{r}_1, t)}{\partial t} = \nabla_1 \cdot \left[\mathbf{D}_1\ \nabla_1 p_1(\mathbf{r}_1, t) \right] \\ - \int_{\partial\mathrm{B}_\sigma(\mathbf{r}_1)} \mathrm{d}\mathcal{S}_2\ \hat{\mathbf{n}}_2 \cdot \left[ \mathbf{D}_1^\text{T} \left( \nabla_1 + \nabla_2 \right) P(\mathbf{r}_1, \mathbf{r}_2, t)\right].
\end{multline}

We are left with evaluating the integral living on the collision surface of the two disks, which we refer to as the collision integral. We denote it by

\begin{equation}
\label{collision_integral}
\mathrm{I}(\mathbf{r}_1, t) = - \int_{\partial\mathrm{B}_\sigma(\mathbf{r}_1)} \mathrm{d}\mathcal{S}_2\ \hat{\mathbf{n}}_2 \cdot \left[ \mathbf{D}_1^\text{T} \left( \nabla_1 + \nabla_2 \right) P(t)\right].
\end{equation}

In the spirit of treating the hard-core collisions between the particles in a geometric sense, we will us a method based on a matched asymptotic expansion to compute $\mathrm{I}(\mathbf{r}_1, t)$ and obtain a closed one-body equation for the time-evolution of $p_1$. 

\subsubsection{Justification of Eg.\eqref{applied_generalized_RTT}}
\label{section_justification_of_Eq_9}

We define a scalar-valued weight function  $f(\mathbf{x}, \mathbf{y})$ for a circular volume $V(\mathbf{x})$ centered around $\mathbf{x}$  in two dimensions as follows

\begin{equation}
\label{th_app_f-definition}
f(\mathbf{x}, \mathbf{y}) \equiv
\begin{cases}
> 0 \qquad &, \mathbf{y}\notin V(\mathbf{x}) \\
= 0 \qquad &, \mathbf{y}\in \partial V(\mathbf{x}) \\
< 0 \qquad &, \mathbf{y}\in V(\mathbf{x})
\end{cases}.
\end{equation}

By this definition, the boundary of the volume can be described as $\partial V(\mathbf{x}) = \{ \mathbf{y}; f(\mathbf{x}, \mathbf{y}) = 0\}$, and the outward normal vector on the boundary is per construction

\begin{equation}
\hat{\mathbf{n}}_\mathbf{y} \equiv \frac{\nabla_\mathbf{y} f(\mathbf{x}, \mathbf{y})}{|\nabla_\mathbf{y} f(\mathbf{x}, \mathbf{y})|}.
\end{equation}

Note that $f(\mathbf{x}, \mathbf{y}) = f(\mathbf{x} - \mathbf{y})$ is only a function of the difference of its variables, such that $\nabla_\mathbf{y} f(\mathbf{x}, \mathbf{y}) = - \nabla_\mathbf{x} f(\mathbf{x}, \mathbf{y})$. This results in 

\begin{equation}
\label{th_app_normal_vectors}
\hat{\mathbf{n}}_\mathbf{x} \equiv \frac{\nabla_\mathbf{x} f(\mathbf{x}, \mathbf{y})}{|\nabla_\mathbf{x} f(\mathbf{x}, \mathbf{y})|} = - \hat{\mathbf{n}}_\mathbf{y},
\end{equation}
valid on $\partial V(\mathbf{x})$. 

Note that in Eq.\eqref{th_app_f-definition} we defined the function $f$ such, that the resulting normal vector is an outward normal vector. Whether $\mathbf{y}$ belongs to $V(\mathbf{x})$ therefore is determined by the characteristic function $\Theta(-f(\mathbf{x}, \mathbf{y}))$, where $\Theta(\cdot)$ denotes the Heaviside-step function.

We are interested in taking the divergence of a vector-valued function $\mathbf{h}(\mathbf{x}, \mathbf{y})$ integrated over the volume $V(\mathbf{x})$. We can manipulate the resulting expression using general properties of the Heaviside-step function and its derivative, the Dirac-delta function $\delta(\cdot)$

\begin{widetext}
\begin{subequations}
\begin{align}
\nabla_\mathbf{x} \cdot \int_{V(\mathbf{x})}\mathrm{d}\mathbf{y}\ \mathbf{h}(\mathbf{x}, \mathbf{y}) &= \nabla_\mathbf{x} \cdot \int_{\mathbb{R}^2}\mathrm{d}\mathbf{y}\ \mathbf{h}(\mathbf{x}, \mathbf{y}) \ \Theta(-f(\mathbf{x}, \mathbf{y}))\\
&= \int_{\mathbb{R}^d}\mathrm{d}\mathbf{y}\ [\nabla_\mathbf{x}\cdot \mathbf{h}(\mathbf{x}, \mathbf{y})]\ \Theta(-f(\mathbf{x}, \mathbf{y})) - \int_{\mathbb{R}^d}\mathrm{d}\mathbf{y}\ \mathbf{h}(\mathbf{x}, \mathbf{y})\cdot [\delta(f(\mathbf{x}, \mathbf{y}))\ \nabla_\mathbf{x}f(\mathbf{x}, \mathbf{y})] \\
&= \int_{V(\mathbf{x})}\mathrm{d}\mathbf{y}\ \nabla_\mathbf{x}\cdot \mathbf{h}(\mathbf{x}, \mathbf{y}) - \int_{\partial V(\mathbf{x})}\mathrm{d}\mathcal{S}_y\ \mathbf{h}(\mathbf{x}, \mathbf{y})\cdot \frac{\nabla_\mathbf{x}f(\mathbf{x}, \mathbf{y})}{|\nabla_\mathbf{x}f(\mathbf{x}, \mathbf{y})|} \\
\label{th_app_intermedeiate_eq}
&= \int_{V(\mathbf{x})}\mathrm{d}\mathbf{y}\ \nabla_\mathbf{x}\cdot \mathbf{h}(\mathbf{x}, \mathbf{y}) - \int_{\partial V(\mathbf{x})}\mathrm{d}\mathcal{S}_y\ \mathbf{h}(\mathbf{x}, \mathbf{y}) \cdot \hat{\mathbf{n}}_\mathbf{x}.
\end{align}
\end{subequations}
\end{widetext}
Here $\mathbf{a} \cdot \mathbf{b}$ denotes the inner product between two vectors $\mathbf{a}$, $\mathbf{b}$ and $\mathrm{d}\mathcal{S}_y$ represents the surface element with respect to $\mathbf{y}$-integration.

Using Eq.\eqref{th_app_normal_vectors} we can replace $\hat{\mathbf{n}}_\mathbf{x} = -\hat{\mathbf{n}}_\mathbf{y}$ on $\partial V(\mathbf{x})$, and rearrange Eq.\eqref{th_app_intermedeiate_eq} to

\begin{multline}
\label{th_app_transport_vector}
\int_{V(\mathbf{x})}\mathrm{d}\mathbf{y}\ \nabla_\mathbf{x}\cdot \mathbf{h}(\mathbf{x}, \mathbf{y}) = \nabla_\mathbf{x}\cdot \int_{V(\mathbf{x})}\mathrm{d}\mathbf{y}\ \mathbf{h}(\mathbf{x}, \mathbf{y}) \\ 
 - \int_{\partial V(\mathbf{x})}\mathrm{d}\mathcal{S}_y\ \mathbf{h}(\mathbf{x}, \mathbf{y})\cdot \hat{\mathbf{n}}_\mathbf{y}.
\end{multline}

To arrive at Eq.\eqref{applied_generalized_RTT}, we need to take a second derivative with respect to $\mathbf{x}$, i.e., we need to evaluate $\nabla_\mathbf{x} \cdot \left[\mathbf{A} \nabla_\mathbf{x} \int_{V(\mathbf{x})}\mathrm{d}\mathbf{y}\ h \right]$, where $\mathbf{A}$ is a space-independent matrix and $\mathbf{h}(\mathbf{x}, \mathbf{y}) = h(\mathbf{x}, \mathbf{y}) \left(\begin{smallmatrix}1 \\ 1\end{smallmatrix}\right)$. Using the result of Eq.\eqref{th_app_transport_vector}, we obtain

\begin{multline}
\nabla_\mathbf{x} \cdot \left[\mathbf{A} \nabla_\mathbf{x} \int_{V(\mathbf{x})}\mathrm{d}\mathbf{y}\ h \right] = \nabla_\mathbf{x} \cdot \left[\int_{V(\mathbf{x})}\mathrm{d}\mathbf{y}\ \mathbf{A} \nabla_\mathbf{x}\ h\right. \\  + \left.\int_{\partial V(\mathbf{x})}\mathrm{d}\mathcal{S}_y\ \mathbf{A} \hat{\mathbf{n}}_\mathbf{y}\ h\right].
\end{multline}

Evaluating the second derivative, we again make use of Eq.\eqref{th_app_transport_vector}

\begin{subequations}
\label{th_app_inbetween_result_1}
\begin{align}
&\nabla_\mathbf{x} \cdot \left[\mathbf{A} \nabla_\mathbf{x} \int_{V(\mathbf{x})}\mathrm{d}\mathbf{y}\ h \right] =  \int_{V(\mathbf{x})}\mathrm{d}\mathbf{y}\ \nabla_\mathbf{x} \cdot\left(\mathbf{A} \nabla_\mathbf{x}\ h\right) \\
&\qquad\quad- \int_{\partial V(\mathbf{x})}\mathrm{d}\mathcal{S}_y\  \hat{\mathbf{n}}_\mathbf{y}\cdot \mathbf{A} \nabla_\mathbf{x}\ h \\
&\qquad\quad + \int_{\partial V(\mathbf{x})}\mathrm{d}\mathcal{S}_y\ \hat{\mathbf{n}}_\mathbf{y} \cdot \mathbf{A}^\text{T}\nabla_\mathbf{x}\ h\\
\label{th_app_inbetween_result_1a}
&\qquad\quad + \int_{\partial V(\mathbf{x})}\mathrm{d}\mathcal{S}_y\ \big[\mathbf{A}\ h\big] : \big[\nabla_\mathbf{x} \otimes \hat{\mathbf{n}}_\mathbf{y}\big], 
\end{align}
\end{subequations}
where $\mathbf{a}\otimes\mathbf{b}$ denotes an outer product between the two vectors $\mathbf{a}$, $\mathbf{b}$ and $\mathbf{A}:\mathbf{B}$ the double contraction between two tensors $\mathbf{A}, \mathbf{B}$. Note that in the last term we can replace $\nabla_\mathbf{x}$ by $-\nabla_\mathbf{y}$ due to Eq.\eqref{th_app_normal_vectors}. We now evaluate this last integral in Eq.\eqref{th_app_inbetween_result_1a} by integration by parts

\begin{subequations}
\begin{align}
&\int_{\partial V(\mathbf{x})}\mathrm{d}\mathcal{S}_y\ \Big[\mathbf{A}\ h(\mathbf{x}, \mathbf{y})\Big] : \left[\nabla_\mathbf{x} \otimes \hat{\mathbf{n}}_\mathbf{y}\right] \nonumber \\
&\quad= \int_{\partial V(\mathbf{x})}\mathrm{d}\mathcal{S}_y\ \Big[\mathbf{A}\ h(\mathbf{x}, \mathbf{y})\Big] : \left[\left(-\nabla_\mathbf{y}\right) \otimes \hat{\mathbf{n}}_\mathbf{y}\right] \\
&\quad= - \int_{\partial V(\mathbf{x})}\mathrm{d}\mathcal{S}_y\ \hat{\mathbf{n}}_\mathbf{y} \cdot \big[\mathbf{A}^\text{T} \left(-\nabla_\mathbf{y}\right) h(\mathbf{x}, \mathbf{y})\big]\\
\label{th_app_inbetween_result_2}
&\quad= \int_{\partial V(\mathbf{x})}\mathrm{d}\mathcal{S}_y\ \hat{\mathbf{n}}_{\mathbf{y}} \cdot \left[\mathbf{A}^\text{T}\nabla_\mathbf{y}\ h(\mathbf{x}, \mathbf{y})\right].
\end{align}
\end{subequations}
Note that there are no boundary terms, since $\partial(\partial V(\mathbf{x})) = 0$. Combining Eq.\eqref{th_app_inbetween_result_1} with Eq.\eqref{th_app_inbetween_result_2} and rearranging we find

\begin{multline}
\label{th_app_second_order_transport}
\int_{V(\mathbf{x})}\mathrm{d}\mathbf{y}\ \nabla_\mathbf{x} \cdot\left[\mathbf{A}\ \nabla_\mathbf{x} h\right] = \nabla_\mathbf{x} \cdot \left[\mathbf{A} \nabla_\mathbf{x} \int_{V(\mathbf{x})}\mathrm{d}\mathbf{y}\ h \right] \\ - \int_{\partial V(\mathbf{x})}\mathrm{d}\mathcal{S}_y\ \hat{\mathbf{n}}_{\mathbf{y}} \cdot \left[\left(\mathbf{A} + \mathbf{A}^\text{T}\right) \nabla_\mathbf{x} h + \mathbf{A}^\text{T} \nabla_\mathbf{y}\ h\right].
\end{multline}

Using Eq.\eqref{th_app_second_order_transport}, we now can justify Eq.(9) of the main derivation. Therefore we replace in Eq.\eqref{th_app_second_order_transport} $h = h(\mathbf{x}, \mathbf{y})$ by $P(\mathbf{r}_1, \mathbf{r}_2, t)$ and $\mathbf{A}$ by the diffusion tensor $\mathbf{D}_1$. The volume $V(\mathbf{x})$ becomes $\Omega(\mathbf{r}_1)$.  Note that we can use the definition of the one-body probability distribution $p_1(\mathbf{r}_1,t) = \int_{\Omega(\mathbf{r}_1)} \mathrm{d}\mathbf{x}_2\ P(\mathbf{r}_1, \mathbf{r}_2, t)$ in Eq.\eqref{th_app_second_order_transport}.

\subsection{Matched asymptotic expansion}

\subsubsection{The inner and outer region}
For evaluation of the collision integral in Eq.~\eqref{collision_integral}, we change perspective to describe the problem by adapting the approach of Refs.~\cite{bruna2012excluded, bruna2012diffusion}. This will respect that the particle interaction is localized near the collision surface $\partial \mathrm{B}_\sigma(\mathbf{r}_1)$. 

When the two particles are far apart from each other $(||\mathbf{r}_1 - \mathbf{r}_2|| \gg \sigma)$, they can be treated as independent. In this so called outer region we define $P^\text{out}(\mathbf{r}_1, \mathbf{r}_2, t) \equiv P(\mathbf{r}_1, \mathbf{r}_2, t)$ and by the independency argument we have that $P^\text{out}(\mathbf{r}_1, \mathbf{r}_2, t) = g_1(\mathbf{r}_1, t) g_2(\mathbf{r}_2, t)$ for some functions $g_1$ and $g_2$. Using the normalization condition on $P(t)$, one can find that indeed $g_1 = p_1 + \mathcal{O}(\sigma^2)$ and $g_2 = p_2 + \mathcal{O}(\sigma^2)$.

\begin{figure}[t]
\includegraphics[width=0.95\columnwidth]{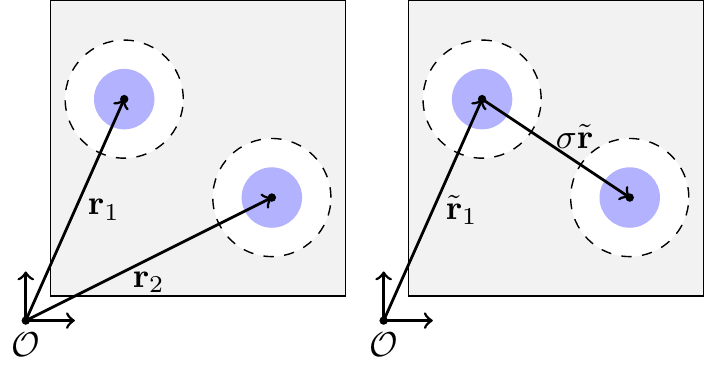}

\caption{When the two particles are close in the inner region, i.e. $|| \mathbf{r}_1 - \mathbf{r}_2|| \sim \sigma$, the set of coordinates describing the positions of the two particles changes from the ordinary (left) to a particle-focused (right). The change of coordinates corresponds to fixing the first particle (previously $\mathbf{r}_1$) and describing the position of the second particle with respect to the position of the first. The advantage of this new set of coordinates for the collision problem is that the collision surface now simply is the area of $||\tilde{\mathbf{r}}|| = 1$. }
\label{fig:inner_coordinates}
\end{figure}

When the two particles are close together $(||\mathbf{r}_1 - \mathbf{r}_2|| \sim \sigma)$, they are correlated and thus, the collision integral will become relevant. In this so called inner region we assign new coordinates to the problem. We consider relative coordinates in which the position of the second particle is obtained with respect to the (given) position of the first particle. Therefore we set $\mathbf{r}_1 \equiv \tilde{\mathbf{r}}_1$ and $\mathbf{r}_2 \equiv \tilde{\mathbf{r}}_1 + \sigma\tilde{\mathbf{r}}$ and define for the inner joint probability $\sigma^2\tilde{P}(\tilde{\mathbf{r}}_1, \tilde{\mathbf{r}}, t) \equiv P(\mathbf{r}_1, \mathbf{r}_2, t)$, to which we assign also a tilde for convenience. Again we use the short notation  $\tilde{P}(t) = \tilde{P}(\tilde{\mathbf{r}}_1, \tilde{\mathbf{r}}, t)$. The coordinate change is illustrated in Fig.~\ref{fig:inner_coordinates}.

The reformulated problem of Eq.~\eqref{general_FP_eq} reads in the new coordinates

\begin{align}
\label{general_FPeq_inner_coordinates}
\sigma^2 \frac{\partial \tilde{P}(t)}{\partial t} &= \nabla_{\tilde{\mathbf{r}}} \cdot \left[(\mathbf{D}_1 + \mathbf{D}_2)\  \nabla_{\tilde{\mathbf{r}}} \tilde{P}(t)\right] \nonumber \\
&\quad -\sigma \nabla_{\tilde{\mathbf{r}}} \cdot \left[ \left(\mathbf{D}_1 + \mathbf{D}_1^\mathrm{T}\right)\nabla_{\tilde{\mathbf{r}}_1} \tilde{P}(t) \right] \nonumber \\
&\quad + \sigma^2 \nabla_{\tilde{\mathbf{r}}_1} \cdot \left[\mathbf{D}_1\  \nabla_{\tilde{\mathbf{r}}_1} \tilde{P}(t)\right],
\end{align}
with an obvious choice for $\nabla_{\tilde{\mathbf{r}}_1}$ and $\nabla_{\tilde{\mathbf{r}}}$ as the partial differential operators with respect to the inner coordinates $\tilde{\mathbf{r}}_1$ and $\tilde{\mathbf{r}}$ respectively. 

In inner coordinates, the collision surface $\partial\mathrm{B}_\sigma(\mathbf{r}_1)$ is the surface $||\tilde{\mathbf{r}}|| = 1$. The boundary condition of Eq.~\eqref{bc_for_general_FP_eq} in inner coordinates therefore reads

\begin{equation}
\label{bc_general_FPeq_inner_coordinates}
\tilde{\mathbf{r}} \cdot \left[(\mathbf{D}_1 + \mathbf{D}_2)\ \nabla_{\tilde{\mathbf{r}}} \tilde{P}(t)\right] = \sigma \tilde{\mathbf{r}} \cdot \left[\mathbf{D}_1 \ \nabla_{\tilde{\mathbf{r}}_1} \tilde{P}(t)\right].
\end{equation}

In inner variables we need to introduce another boundary condition, since the inner solution $\tilde{P}(t)$ has to match the outer solution $P^\text{out}(t)$ as $||\tilde{\mathbf{r}}|| \to \infty$. Taylor-expanding the outer solution in powers of the particle diameter $\sigma$, we obtain

\begin{multline}
\label{new_bc_inner_coordinates}
\tilde{P}(\tilde{\mathbf{r}}_1, \tilde{\mathbf{r}}, t) \sim p_1(\tilde{\mathbf{r}}_1, t)\  p_2(\tilde{\mathbf{r}}_1, t) + \\ + \sigma p_1(\tilde{\mathbf{r}}_1, t)\ \tilde{\mathbf{r}} \cdot \nabla_{\tilde{\mathbf{r}}_1} p_2(\tilde{\mathbf{r}}_1, t) + \mathcal{O}(\sigma^2).
\end{multline}

Equation~\eqref{general_FPeq_inner_coordinates} together with the boundary conditions Eq.~\eqref{bc_general_FPeq_inner_coordinates} (at $||\tilde{\mathbf{r}}|| = 1$) and Eq.~\eqref{new_bc_inner_coordinates} (as $||\tilde{\mathbf{r}}|| \to \infty$) defines the two-particle collision problem in inner coordinates. In the spirit of the matched asymptotic expansion we now look for a perturbative solution of this system in powers of the particle diameter $\sigma$

\begin{equation}
\label{pertubative_ansatz}
\tilde{P}(\tilde{\mathbf{r}}_1, \tilde{\mathbf{r}}, t) = \tilde{P}^{(0)}(\tilde{\mathbf{r}}_1, \tilde{\mathbf{r}}, t) + \sigma \tilde{P}^{(1)}(\tilde{\mathbf{r}}_1, \tilde{\mathbf{r}}, t) + \mathcal{O}(\sigma^2).
\end{equation}

Inserting this ansatz into Eq.~\eqref{general_FPeq_inner_coordinates} yields a Laplace equation for each order, which we refer to as the Laplace problem of the corresponding order. 

\subsubsection{$\tilde{P}^{(0)}$-problem}
The leading order of the Laplace-problem in Eq.~\eqref{general_FPeq_inner_coordinates} with boundary conditions as defined in Eq.~\eqref{bc_general_FPeq_inner_coordinates} and Eq.~\eqref{new_bc_inner_coordinates} reads

\begin{subequations}
\begin{align}
\label{zero_orper_laplace_equation}
\nabla_{\tilde{\mathbf{r}}} \cdot\nabla_{\tilde{\mathbf{r}}} \tilde{P}^{(0)}(t) &= 0, \\
\tilde{\mathbf{r}} \cdot \left[(\mathbf{D}_1 + \mathbf{D}_2)\ \nabla_{\tilde{\mathbf{r}}} \tilde{P}^{(0)}(t)\right] &= 0
\end{align}
on $||\tilde{\mathbf{r}}|| = 1$ and 
\begin{align} 
\tilde{P}^{(0)}(t) &\sim p_1(\tilde{\mathbf{r}}_1, t)\ p_2(\tilde{\mathbf{r}}_1, t)
\end{align}
\end{subequations}
as $||\tilde{\mathbf{r}}|| \to \infty$. Note that due to their isotropic, but anti-symmetric off-diagonal structure the diffusion matrices do not explicitely appear in the Laplace equation \eqref{zero_orper_laplace_equation}. The solution at this order is straight forward to find and is given by

\begin{equation}
\label{P0_solution}
\tilde{P}^{(0)}(\tilde{\mathbf{r}}_1, \tilde{\mathbf{r}}, t) = p_1(\tilde{\mathbf{r}}_1, t)\  p_2(\tilde{\mathbf{r}}_1, t).
\end{equation}

The zero-order problem in the diameter $\sigma$ corresponds to a vanishing collision integral, as it will become apparent later. Hence the problem at this order corresponds to the time-evolution equation of an ideal particle with no hard-core interactions present in the system. 

\subsubsection{$\tilde{P}^{(1)}$-problem}
Including explicitly the obtained solution for $\tilde{P}^{(0)}$ into the first order Laplace problem in Eq.~\eqref{general_FPeq_inner_coordinates} we find 

\begin{multline}
    0 = \nabla_{\tilde{\mathbf{r}}} \cdot \left[(\mathbf{D}_1 + \mathbf{D}_2)\  \nabla_{\tilde{\mathbf{r}}} \tilde{P}^{(1)}(t)\right] \\ - \nabla_{\mathbf{\tilde{r}}} \cdot \underbrace{\left[(\mathbf{D}_1 + \mathbf{D}_1^\mathrm{T})\ \nabla_{\mathbf{\tilde{r}}_1} \tilde{P}^{(0)}(\tilde{\mathbf{r}}_1, t)\right]}_{\mathrm{function\ of}\ \tilde{\mathbf{r}}_1\ \mathrm{only}}.
\end{multline}
Again, due to their specific structure, the diffusion matrices do not appear explicitly in the Laplace equation at first order. But they contribute via the boundary conditions. The first order Laplace problem reads as

\begin{subequations}
\label{P1_problem}
\begin{align}
\label{P1_problem_line1}
\nabla_{\tilde{\mathbf{r}}}\cdot \nabla_{\tilde{\mathbf{r}}} \tilde{P}^{(1)}(t) &= 0, \\
\label{P1_problem_line2}
\tilde{\mathbf{r}} \cdot \left[(\mathbf{D}_1 + \mathbf{D}_2)\ \nabla_{\tilde{\mathbf{r}}} \tilde{P}^{(1)}(t)\right] &= \tilde{\mathbf{r}} \cdot \mathbf{A}(\tilde{\mathbf{r}}_1,t)
\end{align}
on $||\tilde{\mathbf{r}}|| = 1$ and 
\begin{align}
\label{P1_problem_line3}
\tilde{P}^{(1)}(t) &\sim \tilde{\mathbf{r}} \cdot \mathbf{B}(\tilde{\mathbf{r}}_1,t)
\end{align}
\end{subequations}
as $||\tilde{\mathbf{r}}|| \to \infty$. Here

\begin{align}
\mathbf{A}(\tilde{\mathbf{r}}_1,t) &= \mathbf{D}_1\ \nabla_{\tilde{\mathbf{r}}_1} \left(p_1(\tilde{\mathbf{r}}_1, t)\ p_2(\tilde{\mathbf{r}}_1, t)\right), \\
\mathbf{B}(\tilde{\mathbf{r}}_1,t) &= p_1(\tilde{\mathbf{r}}_1, t)\nabla_{\tilde{\mathbf{r}}_1} p_2(\tilde{\mathbf{r}}_1, t).
\end{align}

It is at this point where we essentially include the specific nature of the odd-diffusive system into the excluded-area interactions between the particle, that is the tensorial diffusion with off-diagonal anti-symmetric elements. The solution to the formulated first-order problem in Eqs.~\eqref{P1_problem_line1}-\eqref{P1_problem_line3} reads

\begin{multline}
\label{P1_solution}
\tilde{P}^{(1)}(\tilde{\mathbf{r}}_1, \tilde{\mathbf{r}}, t) = \tilde{\mathbf{r}} \cdot \mathbf{B}(\tilde{\mathbf{r}}_1,t)  + \left[\frac{(\mathbf{D}_1 + \mathbf{D}_2)^\text{T}}{\text{det}(\mathbf{D}_1 + \mathbf{D}_2)} \frac{\tilde{\mathbf{r}}}{||\tilde{\mathbf{r}}||^2}\right] \\ \cdot \left[ (\mathbf{D}_1 + \mathbf{D}_2)\ \mathbf{B}(\tilde{\mathbf{r}}_1,t) - \mathbf{A}(\tilde{\mathbf{r}}_1,t) \right].
\end{multline}
This solution is in agreement with the reported solution for ordinary diffusive systems \cite{bruna2012diffusion}, i.e. $\kappa_1 = \kappa_2 = 0$.

\subsubsection{The collision integral}
We want to use the perturbative ansatz of the matched asymptotic expansion $\tilde{P} = \tilde{P}^{(0)} + \sigma \tilde{P}^{(1)}$ together with the obtained solutions for $\tilde{P}^{(0)}$ and $\tilde{P}^{(1)}$ in Eq.~\eqref{P0_solution} and Eq.~\eqref{P1_solution} to evaluate the collision integral as defined in Eq.~\eqref{collision_integral}. In inner coordinates, the integral reads

\begin{equation}
\label{collision_integral_inner_coordinates}
\mathrm{I}(\tilde{\mathbf{r}}_1, t) = + \sigma \int_{||\tilde{\mathbf{r}}|| = 1} \mathrm{d}\tilde{\mathcal{S}}\ \tilde{\mathbf{r}} \cdot \left[ \mathbf{D}_1^\text{T}\ \nabla_{\tilde{\mathbf{r}}_1} \tilde{P}(t)\right],
\end{equation}
where we used that on the collision surface $||\tilde{\mathbf{r}}|| = 1$ we have that $\hat{\mathbf{n}}_2 = - \hat{\mathbf{n}}_1 = - \tilde{\mathbf{r}}$ and hence $\mathrm{d}\mathcal{S}_2\ \hat{\mathbf{n}}_2 = -\sigma\ \mathrm{d}\tilde{\mathcal{S}}\ \tilde{\mathbf{r}}$. 

Inserting the (truncated) inner solution $\tilde{P}$, the collision integral reads

\begin{widetext}
\begin{subequations}
\begin{align}
\label{zero_order_integral}
\mathrm{I}(\tilde{\mathbf{r}}_1, t) &= \sigma\ \left[\mathbf{D}_1^\text{T} \nabla_{\tilde{\mathbf{r}}_1} (p_1(\tilde{\mathbf{r}}_1, t)\ p_2(\tilde{\mathbf{r}}_1, t))\right] \cdot \int_{||\tilde{\mathbf{r}}|| = 1} \mathrm{d}\tilde{\mathcal{S}}\ \tilde{\mathbf{r}} \\
\label{first_order_integral_1}
&\quad + \sigma^2\ \left[\mathbf{D}_1^\text{T} \nabla_{\tilde{\mathbf{r}}_1} \otimes (p_1(\tilde{\mathbf{r}}_1, t) \nabla_{\tilde{\mathbf{r}}_1} p_2(\tilde{\mathbf{r}}_1, t))\right] : \int_{||\tilde{\mathbf{r}}|| = 1} \mathrm{d}\tilde{\mathcal{S}}\ \tilde{\mathbf{r}} \otimes \tilde{\mathbf{r}} \\
&\quad + \sigma^2 \int_{||\tilde{\mathbf{r}}|| = 1} \mathrm{d}\tilde{\mathcal{S}}\ \tilde{\mathbf{r}} \cdot \mathbf{D}_1^\text{T} \nabla_{\tilde{\mathbf{r}}_1} \left(\left[\frac{(\mathbf{D}_1 + \mathbf{D}_2)^\text{T}}{\text{det}(\mathbf{D}_1 + \mathbf{D}_2)} \tilde{\mathbf{r}}\right] \cdot \left[\mathbf{D}_2\ p_1(\tilde{\mathbf{r}}_1, t) \nabla_{\tilde{\mathbf{r}}_1} p_2(\tilde{\mathbf{r}}_1, t)\right]\right) \\
\label{first_order_integral_3}
&\quad - \sigma^2 \int_{||\tilde{\mathbf{r}}|| = 1} \mathrm{d}\tilde{\mathcal{S}}\ \tilde{\mathbf{r}} \cdot \mathbf{D}_1^\text{T} \nabla_{\tilde{\mathbf{r}}_1} \left(\left[\frac{(\mathbf{D}_1 + \mathbf{D}_2)^\text{T}}{\text{det}(\mathbf{D}_1 + \mathbf{D}_2)} \tilde{\mathbf{r}}\right] \cdot \left[\mathbf{D}_1\ p_2(\tilde{\mathbf{r}}_1, t) \nabla_{\tilde{\mathbf{r}}_1} p_1(\tilde{\mathbf{r}}_1, t) \right]\right).
\end{align}
\end{subequations}
\end{widetext}
Note that $\frac{1}{||\tilde{\mathbf{r}}||} = 1$ on the collision surface $||\tilde{\mathbf{r}}|| = 1$.

In the first integral in Eq.~\eqref{zero_order_integral}, corresponding to the $\tilde{P}^{(0)}$-contribution to the collision integral, we are left with performing an integral of the outward normal vector on the whole unit sphere. With this geometrical insight this integral is exactly zero. Hence the zero-order solution does not contribute to the collision integral as one would expect, since it corresponds to ideal point particles. For the remaining three integrals in Eqs.~\eqref{first_order_integral_1}-\eqref{first_order_integral_3}, the following result is used, which is valid in two dimensions

\begin{equation}
\int_{||\tilde{\mathbf{r}}|| = 1} \mathrm{d}\tilde{\mathcal{S}}\ \tilde{\mathbf{r}} \otimes \tilde{\mathbf{r}} = \pi \mathbf{1}.
\end{equation}

Defining 

\begin{equation}
\boldsymbol{\Lambda}_1 \equiv \mathbf{1} + \frac{\mathbf{D}_1 + \mathbf{D}_2}{\text{det}(\mathbf{D}_1 + \mathbf{D}_2)} \mathbf{D}_2
\end{equation}
and

\begin{equation}
\boldsymbol{\Gamma}_1 \equiv \frac{\mathbf{D}_1 + \mathbf{D}_2}{\text{det}(\mathbf{D}_1 + \mathbf{D}_2)} \mathbf{D}_1,
\end{equation}
the evaluated collision integral reads

\begin{multline}
\label{evaluated_collision_integral_inner_coord}
\mathrm{I}(\tilde{\mathbf{r}}_1, t) = \pi \sigma^2\ \nabla_{\tilde{\mathbf{r}}_1} \cdot \mathbf{D}_1 \left[\boldsymbol{\Lambda}_1\ p_1(\tilde{\mathbf{r}}_1, t) \nabla_{\tilde{\mathbf{r}}_1} p_2(\tilde{\mathbf{r}}_1, t)\right. - \\ - \left.\boldsymbol{\Gamma}_1\ p_2(\tilde{\mathbf{r}}_1, t) \nabla_{\tilde{\mathbf{r}}_1} p_1(\tilde{\mathbf{r}}_1, t)\right]. 
\end{multline}

The back transformation into the original coordinates $\mathbf{r}_1$ and $\mathbf{r}_2$ is straightforward, since the variable $\tilde{\mathbf{r}}$,  representing the separation of the two particles, does not appear anymore in Eq.\eqref{evaluated_collision_integral_inner_coord}. This is a self-consistency check, since we integrated out the effect of the second particle on the first one. With this we have expressed the one-body equation for particle one, as desired, solely in terms of one-body distributions.

\subsection{single-species model}

The presented model captures the effect of two odd-diffusive particles of different species interacting hard-core with each other. The two particles can differ in odd-diffusive parameter and friction, hence their bare diffusivity. But we can also use the presented model to derive the time-evolution equation for the one-body density of two particles of the same species, i.e. $p_1 = p_2 \equiv p$. We still consider odd-diffusive particles, but their diffusivity of course is described by the same tensor $\mathbf{D}_1 = \mathbf{D}_2 \equiv \mathbf{D}$. The time-evolution equation for $p$ can be written straightforwardly from Eq.~\eqref{p1_equation_with_collision_integral} to be

\begin{multline}
\label{p_equation_with_collision_integral}
\frac{\partial p(\mathbf{r}_1, t)}{\partial t} = \nabla_1 \cdot \left[\mathbf{D}\ \nabla_1 p(\mathbf{r}_1, t) \right] \\ - \int_{\partial\mathrm{B}_\sigma(\mathbf{r}_1)} \mathrm{d}\mathcal{S}_2\ \hat{\mathbf{n}}_2 \cdot \left[ \mathbf{D}^\text{T} \left( \nabla_1 + \nabla_2 \right) P(\mathbf{r}_1, \mathbf{r}_2, t)\right].
\end{multline}

The collision integral can be rewritten into inner coordinates in a similar way as in the two-species model, but reduces significantly due to the absence of the inter-species effects. Equation~\eqref{evaluated_collision_integral_inner_coord} reads now

\begin{equation}
\mathrm{I}(\tilde{\mathbf{r}}_1, t) = \pi \sigma^2\ \nabla_{\tilde{\mathbf{r}}_1} \cdot \mathbf{D}\left[ p(\tilde{\mathbf{r}}_1, t) \nabla_{\tilde{\mathbf{r}}_1} p(\tilde{\mathbf{r}}_1, t)\right].
\end{equation} 

As one would have expected, the variable $\tilde{\mathbf{r}}$, as representing the second particle, vanishes at this order and we can simply transform back into the original variables $\mathbf{r}_1$ and $\mathbf{r}_2$. The closed equation on the one-body density level reads

\begin{align}
\label{p_equation}
\frac{\partial p(\mathbf{r}_1, t)}{\partial t} &= \nabla_1 \cdot \mathbf{D}\left[\nabla_1 p(\mathbf{r}_1, t) \right. \nonumber \\ &\quad + \left.\pi \sigma^2\ p(\mathbf{r}_1, t) \nabla_1 p(\mathbf{r}_1, t)\right].
\end{align}

The extension from two particles in the system to an arbitrary number of particles $N$ can be implemented straightforwardly at the order of $\sigma^2$. Since at this order only pairwise interactions are relevant, we know that one particle has $N-1$ inner regions, one with each of the other particles. Thus we can write

\begin{align}
\label{p_equation_final}
\frac{\partial p(\mathbf{r}_1, t)}{\partial t} &= \nabla_1 \cdot \mathbf{D}\left[\nabla_1 p(\mathbf{r}_1, t) \right. \nonumber\\ & \quad + \left.(N-1)\pi \sigma^2\ p(\mathbf{r}_1, t) \nabla_1 p(\mathbf{r}_1, t)\right].
\end{align}

It is apparent in this equation, that hard-core interactions lead to a modified diffusion. The additional term is proportional to the excluded area $\frac{\pi \sigma^2}{4}$ in two dimensions. Eq.~\eqref{p_equation_final} is consistent with the excluded-area modification for ordinary diffusive systems \cite{bruna2012excluded, bruna2012diffusion} and also earlier work, where $p$ was assumed to be close to equilibrium \cite{felderhof1978diffusion}, an assumption we do not need.

\subsection{final equation}

The necessity of the detour to the single-species system will become apparent shortly. Now we are concerned with including the result of the evaluated collision integral in  Eq.~\eqref{evaluated_collision_integral_inner_coord} into the defining one-body equation in  Eq.~\eqref{p1_equation_with_collision_integral} for the density of particle one. Since we arrived at an equation on the one-body level, we only need one variable to describe the motion. Therefore we abbreviate $p_1 \equiv p_1(\mathbf{r}, t)$ and $p_2 \equiv p_2(\mathbf{r}, t)$. Furthermore we denote partial derivatives with respect to this coordinate $\mathbf{r}$ simply as $\nabla$. The time-evolution equation for the one-body density $p_1$ thus reads

\begin{align}
\label{p1_equation}
\frac{\partial p_1(\mathbf{r}, t)}{\partial t} &= \nabla \cdot \mathbf{D}_1\left[\nabla p_1 \right. \nonumber \\ &\quad + \left. \pi \sigma^2\ \left(\boldsymbol{\Lambda}_1\ p_1 \nabla p_2 - \boldsymbol{\Gamma}_1\ p_2 \nabla p_1\right)\right].
\end{align}

The extension of one to $N_2$ particles of species two is straightforward up to order $\sigma^2$, since at this order only pairwise particle interactions need to be considered. For an arbitrary $N_2$, the particle of species one has $N_2$ inner regions with particles of species two, one with each of the $N_2$ particles. Hence there are $N_2$ copies of the corresponding non-linear term in Eq.~\eqref{p1_equation}. The same holds for each of the $N_1$ particles of species one. The particle in focus can have $N_1-1$ pairwise interactions with the remaining particles of species one. As yet the model does not capture these intra-species contributions, we have to recall the non-linear term from the single-species model in Eq.~\eqref{p_equation_final}. The time-evolution equation for the one-body distribution of a particle of species one on a level of pairwise-particle interactions thus reads

\begin{multline}
\label{p1_equation_final}
\frac{\partial p_1(\mathbf{r}, t)}{\partial t} = \nabla \cdot \mathbf{D}_1\left[\nabla p_1 + (N_1 -1)\sigma^2\pi\ p_1 \nabla p_1 \right. + \\ + \left. N_2 \pi \sigma^2\ \left(\boldsymbol{\Lambda}_1\ p_1 \nabla p_2 - \boldsymbol{\Gamma}_1\ p_2 \nabla p_1\right) \right].
\end{multline}

Alternatively to the presented route, we also could have derived the time-evolution equation for $p_2$. But this now can also be read from Eq.~\eqref{p1_equation_final} by an interchange of particle labels to be 

\begin{multline}
\label{p2_equation_final}
\frac{\partial p_2(\mathbf{r}, t)}{\partial t} = \nabla \cdot \mathbf{D}_2\left[\nabla p_2 + (N_2 -1)\sigma^2\pi\ p_2 \nabla p_2 \right. + \\ + \left. N_1 \pi \sigma^2\ \left(\boldsymbol{\Lambda}_2\ p_2 \nabla p_1 - \boldsymbol{\Gamma}_2\ p_1 \nabla p_2\right) \right].
\end{multline}

The matrices in Eq.~\eqref{p1_equation_final} and Eq.~\eqref{p2_equation_final} are ($i = 1, j = 2$ and vice versa)

\begin{align}
\boldsymbol{\Lambda}_i &\equiv \mathbf{1} + \frac{\mathbf{D}_i + \mathbf{D}_j}{\text{det}(\mathbf{D}_i + \mathbf{D}_j)} \mathbf{D}_j,\\
\boldsymbol{\Gamma}_i &\equiv \frac{\mathbf{D}_i + \mathbf{D}_j}{\text{det}(\mathbf{D}_i + \mathbf{D}_j)} \mathbf{D}_i.
\end{align}

Equation~\eqref{p1_equation_final} and Eq.~\eqref{p2_equation_final} together may be written in a joint matrix form

\begin{equation}
\label{p1_p2_matrix_eq}
\frac{\partial}{\partial t} \begin{pmatrix}p_1(\mathbf{r},t) \\ p_2(\mathbf{r},t)\end{pmatrix} = \begin{pmatrix}\nabla\\ \nabla \end{pmatrix} \cdot \left[\mathbb{D}(p_1, p_2) \begin{pmatrix}
\nabla p_1(\mathbf{r},t) \\ \nabla p_2(\mathbf{r},t) \end{pmatrix}\right],
\end{equation}

where 

\begin{widetext}
\begin{equation}
\mathbb{D}(p_1,p_2) = \begin{bmatrix}
\mathbf{D}_1 \left(1 + (N_1 - 1) \sigma^2 \pi\ p_1 - N_2 \sigma^2 \pi\  \boldsymbol{\Gamma}_2\ p_2\right) & N_2 \mathbf{D}_1 \sigma^2 \pi\ \boldsymbol{\Lambda}_1\ p_1 \\ N_1 \mathbf{D}_2 \sigma^2 \pi\ \boldsymbol{\Lambda}_2\ p_2 & \mathbf{D}_2 \left(1 + (N_2 - 1) \sigma^2 \pi\ p_2 - N_1 \sigma^2 \pi \boldsymbol{\Gamma}_1\ p_1\right) \end{bmatrix}
\end{equation}
\end{widetext}
 is the diffusion matrix, both depending on space and time via $p_1(\mathbf{r}, t)$ and $p_2(\mathbf{r},t)$.
 
\section{Outline of First Principles approach}
\label{section_outline}

The asymptotic approach presented in our manuscript is more general and applicable to nonequilibrium odd-diffusive systems, such as active chiral particles. Despite the fact that simulations confirmed our model predictions of interactions-enhanced self-diffusion in odd-diffusive systems, we decided to take an alternative approach in order to gain physical insight into the predicted phenomenon. It is this consideration that motivated us to take a first-principles approach based on the force autocorrelation function (FACF) which is firmly rooted in statistical mechanics to investigate self-diffusion. Charged Brownian particles under magnetic field constitute an equilibrium odd-diffusive system. Moreover, there is a vast body of literature on the FACF of hard spheres with exact analytical results in the low density limit. Below we present a survey of the essential steps in the derivation of the FACF. Our goal is to show the reader that the physical mechanism of enhancement, namely the mutual rolling effect can be inferred from the temporal behavior of the FACF.\\

Our approach below closely follows the original work of Hanna, Hess and Klein \cite{hanna1981velocity, hanna1982self}.
The time evolution equation for the joint probability distribution $P(t) = P(\mathbf{r}_1, \mathbf{r}_2, t)$ of two Brownian particles with position vectors $\rr_1$ and $\rr_2$ can be written as

\begin{align}
\label{FACF_smoluchowski_eq}
\frac{\partial }{\partial t} P(t) &= \nabla_1 \cdot \mathbf{D} \left[\nabla_1 + \beta\ \nabla_1 U(\mathbf{r}_1, \mathbf{r}_2)\right] P(t) \nonumber \\
&\quad+ \nabla_2 \cdot \mathbf{D} \left[\nabla_2 + \beta\ \nabla_2 U(\mathbf{r}_1, \mathbf{r}_2)\right] P(t),
\end{align}
where $\nabla_i$ denotes the partial differentiation with respect to particle coordinate $\mathbf{r}_i$ for $i \in \{1,2\}$ and $\beta$ is the inverse temperature. The particles are equal sized with diameter $\sigma$ and interact with each other via a hard-core interaction potential $U(\mathbf{r}_1, \mathbf{r}_2)$. The diffusive behavior of both particles is described by the tensor $\mathbf{D} = D_0\left(\begin{smallmatrix} 1 & \kappa \\ -\kappa & 1\end{smallmatrix}\right)$. Note that the tensor describes isotropic diffusion with bare diffusivity $D_0$. 

Eq.\eqref{FACF_smoluchowski_eq} decouples into independent equations for the center of mass coordinate $\mathbf{R} = (\rr_1 + \rr_2)/2$ and the inner coordinate $\mathbf{r} = (\rr_1 - \rr_2)$ as
\begin{align}
\label{FACF_equation_for_com_coordinate}
    \frac{\partial}{\partial t} \rho(\mathbf{R}, t) &= \frac{1}{2} \nabla_{\mathbf{R}} \cdot \left[ D_0\ \nabla_{\mathbf{R}}\right] \rho(\mathbf{R}, t), \\
    \label{FACF_equation_for_inner_coordinate}
    \frac{\partial}{\partial t} \varrho(\mathbf{r}, t) &= 2 \nabla_\mathbf{r} \cdot \left[D_0\ \mathbf{1}\ \nabla_\mathbf{r} + \beta\ \mathbf{D}\ \nabla_\mathbf{r} U(\mathbf{r})\right] \varrho(\mathbf{r}, t), 
\end{align}
where $\nabla_\mathbf{R}, \nabla_\mathbf{r}$ denote partial derivatives with respect to the coordinates $\mathbf{R}, \mathbf{r}$. $\rho(\mathbf{R}, t)$ and $\varrho(\mathbf{r}, t)$ denote the distributions of the center of mass coordinate and the inner coordinate, respectively. The hard-core interaction potential takes a simple form in terms of the inner coordinate $\mathbf{r} = r \hat{\mathbf{r}}$ as $U(r) = \left\{\begin{smallmatrix}0, & r > \sigma \\ \infty, & r \leq \sigma \end{smallmatrix}\right.$. Apparently the characteristic tensorial diffusion in an odd-diffusive system only affects the dynamics of the inner coordinate (Eq.\eqref{FACF_equation_for_inner_coordinate}).
The equation for the center of mass coordinate admits a Gaussian solution with a (scalar) diffusion coefficient $D_0/2$.

We are interested in solving for the conditional distribution of the inner coordinate $\varrho(\mathbf{r},t|\mathbf{r}_0) = \varrho(\mathbf{r},t|\mathbf{r}_0,t_0= 0)$, which is the probability distribution of finding the two particles at a distance $\mathbf{r}$ from each other at time $t$, given that initially at $t_0 = 0$ they were at a distance $|\mathbf{r}_0| > \sigma$. The solution can be found using standard tools for partial differential equations and reads in the Laplace-domain as follows

\begin{widetext}
\begin{equation}
\label{FACF_out_inner_solution}
\tilde{\varrho}(\mathbf{r},s|\mathbf{r}_0) = \Theta(r - \sigma)\ \sum_{n=-\infty}^\infty  \frac{\mathrm{e}^{\mathrm{i}n(\varphi - \varphi_0)}}{2\pi}\ \frac{\sigma^2}{2 D_0} \int_0^\infty \mathrm{d}u\ u\frac{J_n(u r_0)}{z^2 + u^2 \sigma^2} \left[J_n(ur) - K_n\left(\frac{zr}{\sigma}\right)\frac{u\sigma\ J_n'(u\sigma) + \mathrm{i}n\kappa\ J_n(u\sigma)}{z\ K_n'(z) + \mathrm{i}n \kappa\ K_n(z)}\right].
\end{equation}
\end{widetext}
Here $\mathrm{i}$ denotes the imaginary unit, $u$ is a dummy integration variable and we introduced $z$ for a shortness of notation as a modified Laplace-variable $z = \sqrt{\frac{s\sigma^2}{2 D_0}}$, $s$ denotes the original Laplace variable. The inner coordinate is decomposed into its polar representation in two dimensions according to $\mathbf{r} = (r,\varphi)$, similarly $\mathbf{r}_0 = (r_0, \varphi_0)$. $J_n(x)$ is the Bessel function of first kind and $K_n(x)$ the modified Bessel function of third kind, both of order $n$. We have used the notation $f'(c) = \frac{\mathrm{d} f(x)}{\mathrm{d} x} \big|_{x = c}$ as a shorthand notation of a derivative of a function $f$ evaluated at $x=c$. Note that in Eq.\eqref{FACF_out_inner_solution} the Heaviside step-function $\Theta(r - \sigma)$ ensures the excluded volume condition present in the hard-sphere system.

The FACF is given as an ensemble average of the force $\mathbf{F} = -\nabla_\mathbf{r} U(r)$ at time $t$ projected on the force at initial time $t=0$, formally $C_F(t) = \frac{1}{2}\langle \mathbf{F}(t) \cdot \mathbf{F}(0)\rangle$. Using established tools for hard-sphere systems, one can rewrite the singular inter-particle collision forces appearing in the definition of $C_F(t)$ by using the solution of the conditional probability of the inner coordinate $\varrho$. In the dilute limit, i.e., considering up to two-body collisions, $C_F(s)$ can be written in the Laplace-domain as

\begin{align}
\tilde{C}_F(s) &= \frac{2}{\beta^2} \frac{N}{V} \int\mathrm{d}\mathbf{r}\int\mathbf{r}_0\ \delta(r - \sigma)\ \delta(r_0 - \sigma) \nonumber\\ &\quad \times \tilde{\varrho}(\mathbf{r}, s|\mathbf{r}_0)\ \hat{\mathbf{r}}\cdot\hat{\mathbf{\mathbf{r}}}_0,
\end{align}
where the conditional distribution of the inner coordinate $\tilde{\varrho}$ of Eq.\eqref{FACF_out_inner_solution} appears explicitly. Note that the FACF only contains contributions from the (unit) radial components $\hat{\mathbf{r}}\cdot\hat{\mathbf{r}}_0$, specifically when $r = \sigma$, i.e., when the two particles are in contact with each other. The Dirac delta functions $\delta(r - \sigma)$ and $\delta (r_0  -\sigma)$ ensure that the singular interaction potential of hard-spheres only contributes at contact distance $r = \sigma, r_0 = \sigma$. 

\begin{figure}[t]
\centering
\includegraphics[width=\columnwidth]{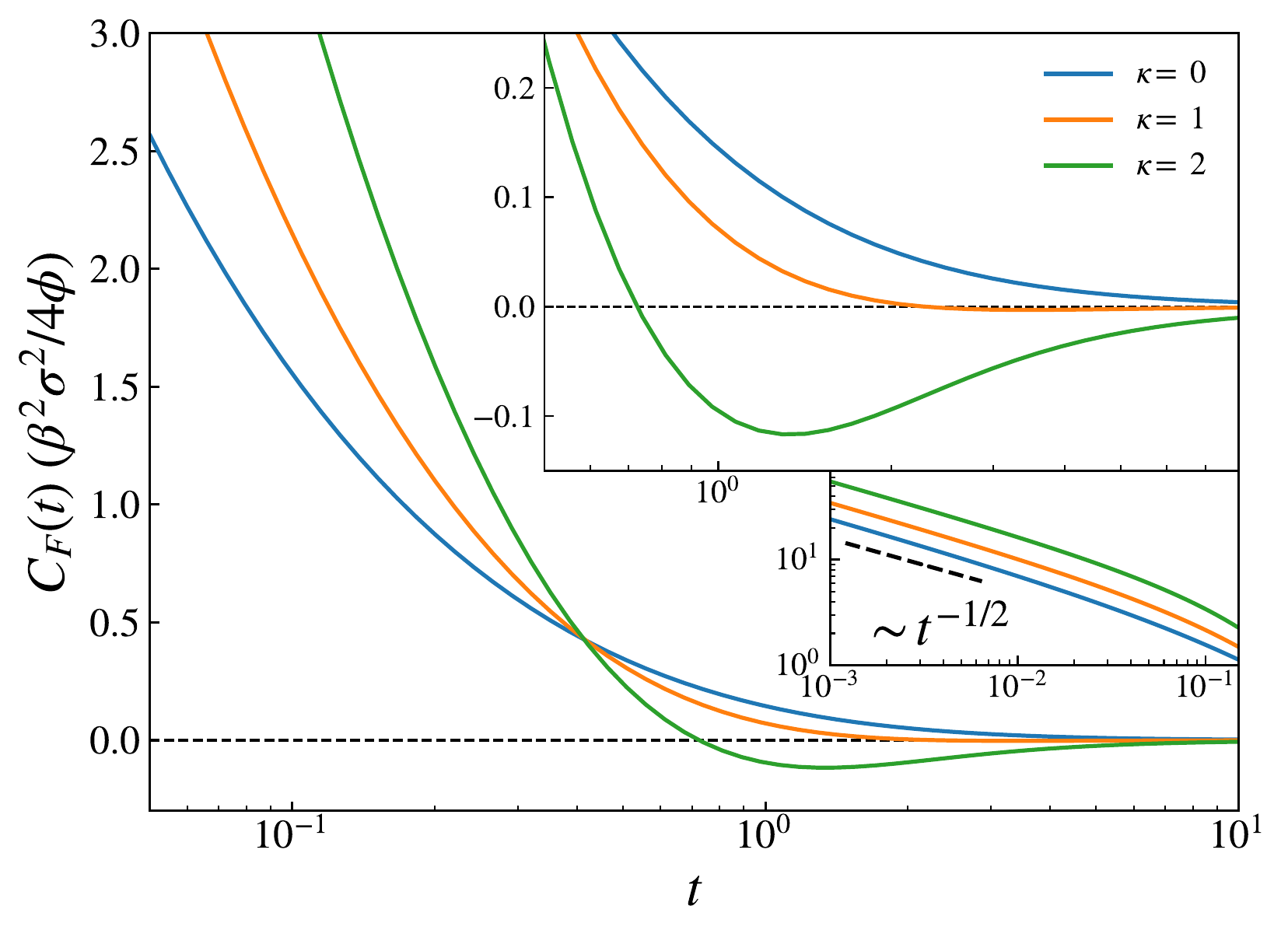}
\caption{Force autocorrelation function (FACF) $C_F(t) = \frac{1}{2}\langle \mathbf{F}(t) \cdot \mathbf{F}(0)\rangle$ for the equilibrium odd-diffusive system of charged Brownian particles subjected to Lorentz force. When particles are normal diffusing ($\kappa =0$), the FACF is a positive monotonically decaying function of time. For odd-diffusive particles, however, the FACF exhibits zero crossing in time. This is consistent with the collision induced mutual rolling effect to explain the enhanced self-diffusion. Due to the singular nature of interaction, the FACF diverges in zero time limit as $t^{-1/2}$ [4] for all values of $\kappa$.}
\label{fig:facf}
\end{figure}

Using the solution of the inner conditional probability density $\tilde{\varrho}$, we can evaluate $\tilde{C}_F(s)$ and numerically invert it back into real time. The result is plotted for different $\kappa$ in Fig.~\ref{fig:facf}. We see that with increasing $\kappa$, $C_F(t)$ turns negative in time. That the FACF turns negative in time is consistent with the physical picture of particles mutually rolling around each other. With increasing $\kappa$, the effect becomes more pronounced; there is a reversal in the direction of force that the particle experiences due to collisions and these negative correlations persist over increasing times with increasing odd-diffusivity. This behaviour can be seen in Fig.~\ref{fig:facf} (see especially the top inset for the negative turning at $\kappa\neq 0$). Previous work on the FACF of hard-sphere systems \cite{hanna1981velocity} already showed that due to the singular nature of interaction, the FACF diverges in zero time limit as $t^{-1/2}$, which we also confirm for odd-diffusive systems (see bottom inset in Fig.~\ref{fig:facf}).

\begin{figure*}
\includegraphics[width=0.48\textwidth]{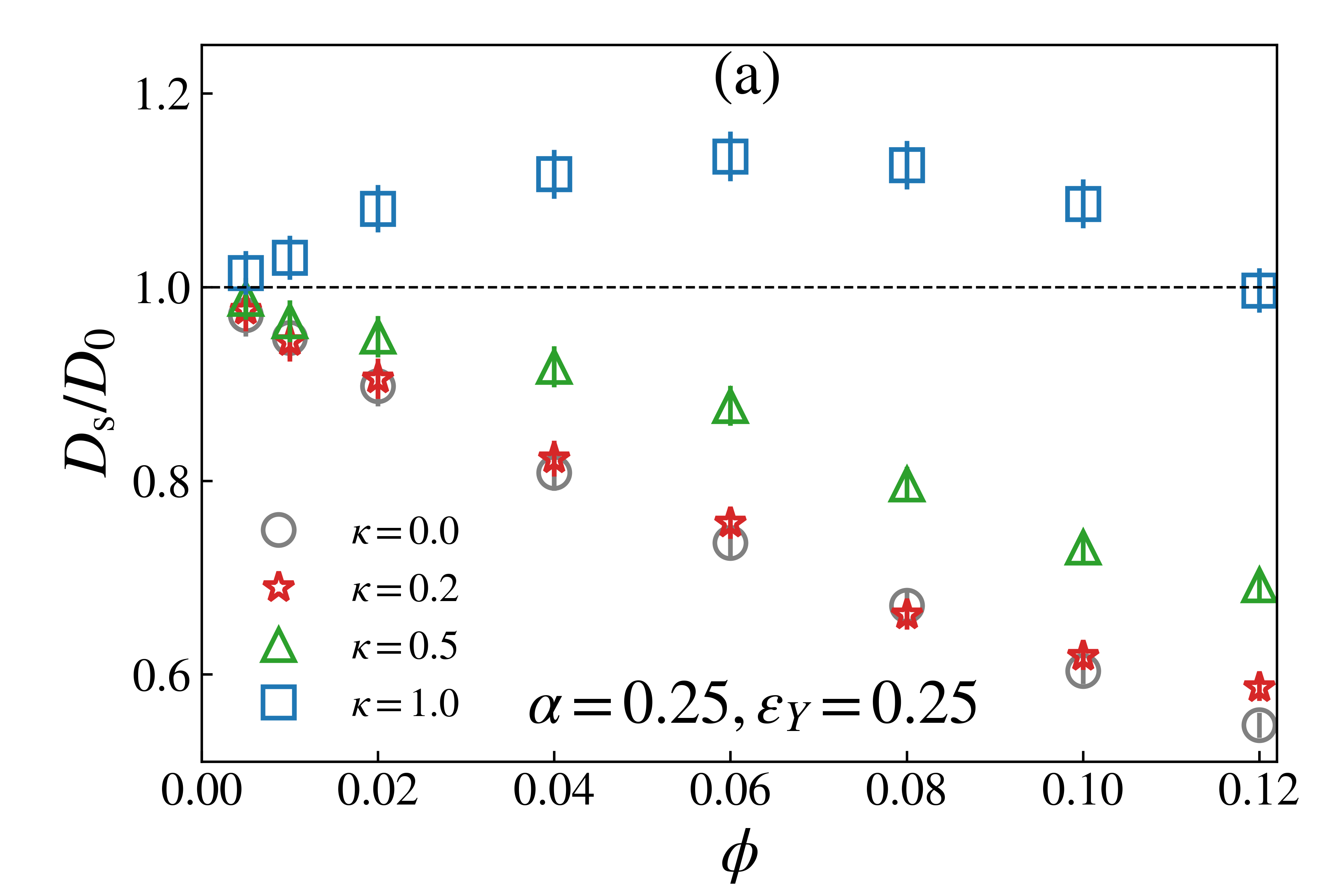}
\includegraphics[width=0.48\textwidth]{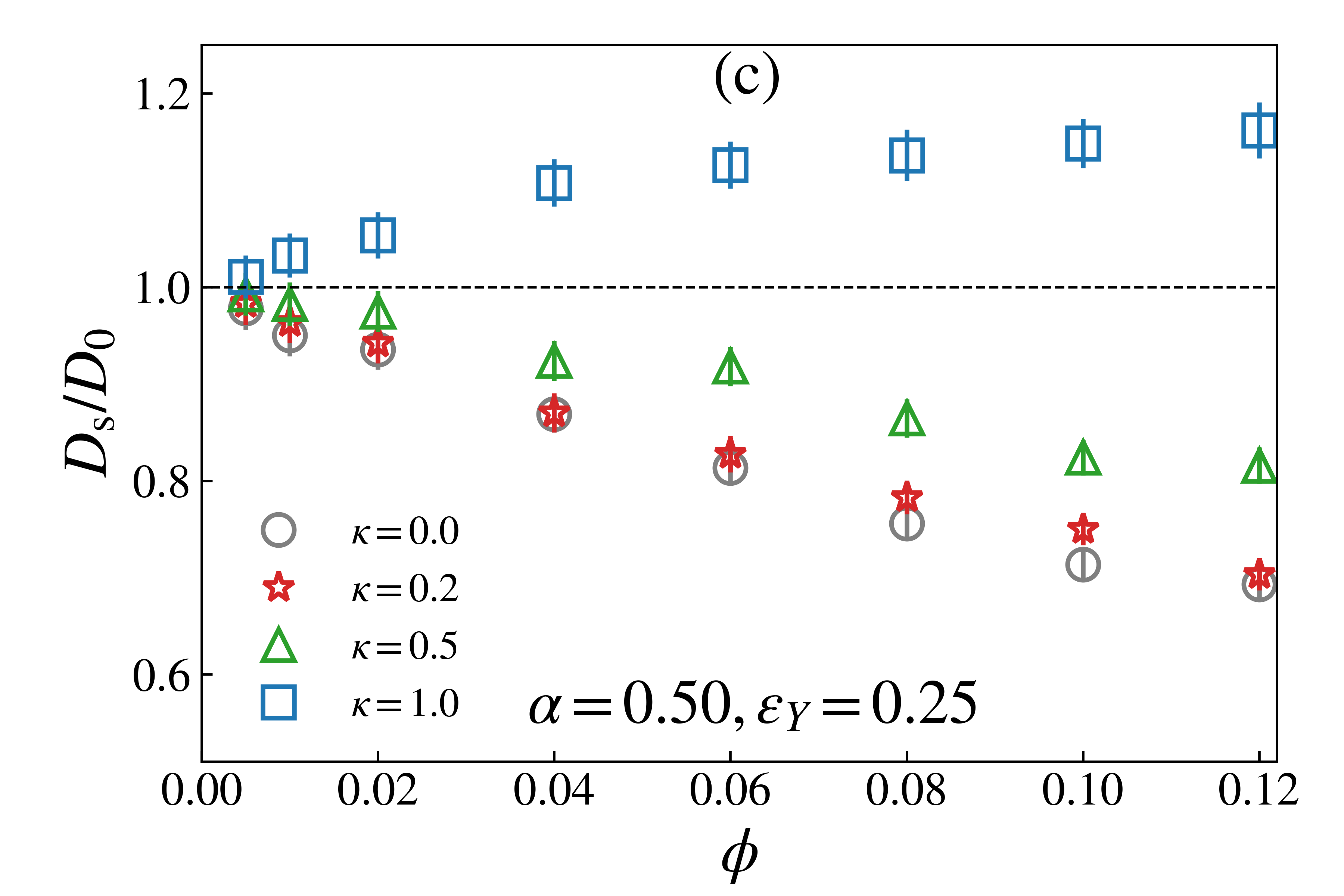}
\includegraphics[width=0.48\textwidth]{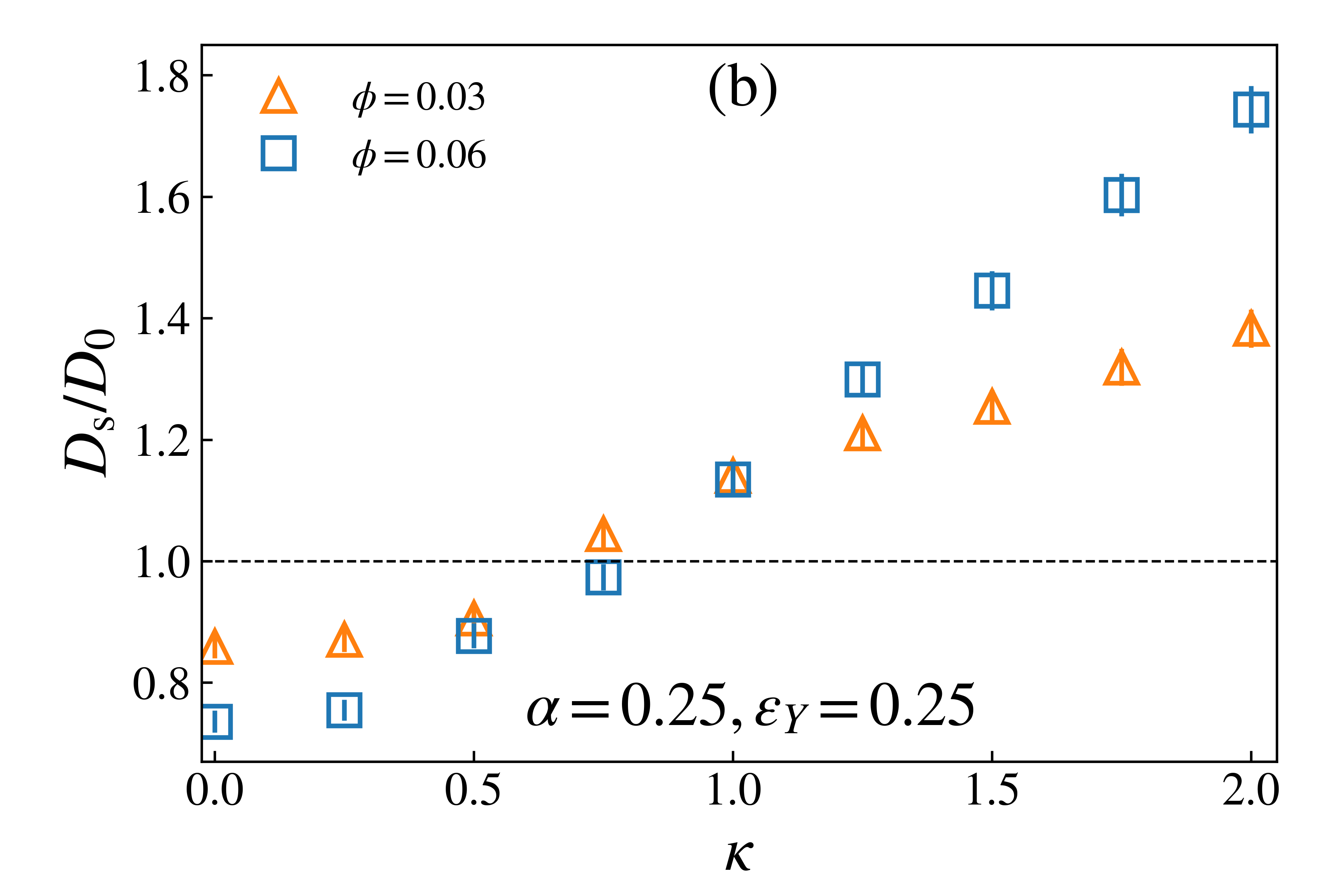}
\includegraphics[width=0.48\textwidth]{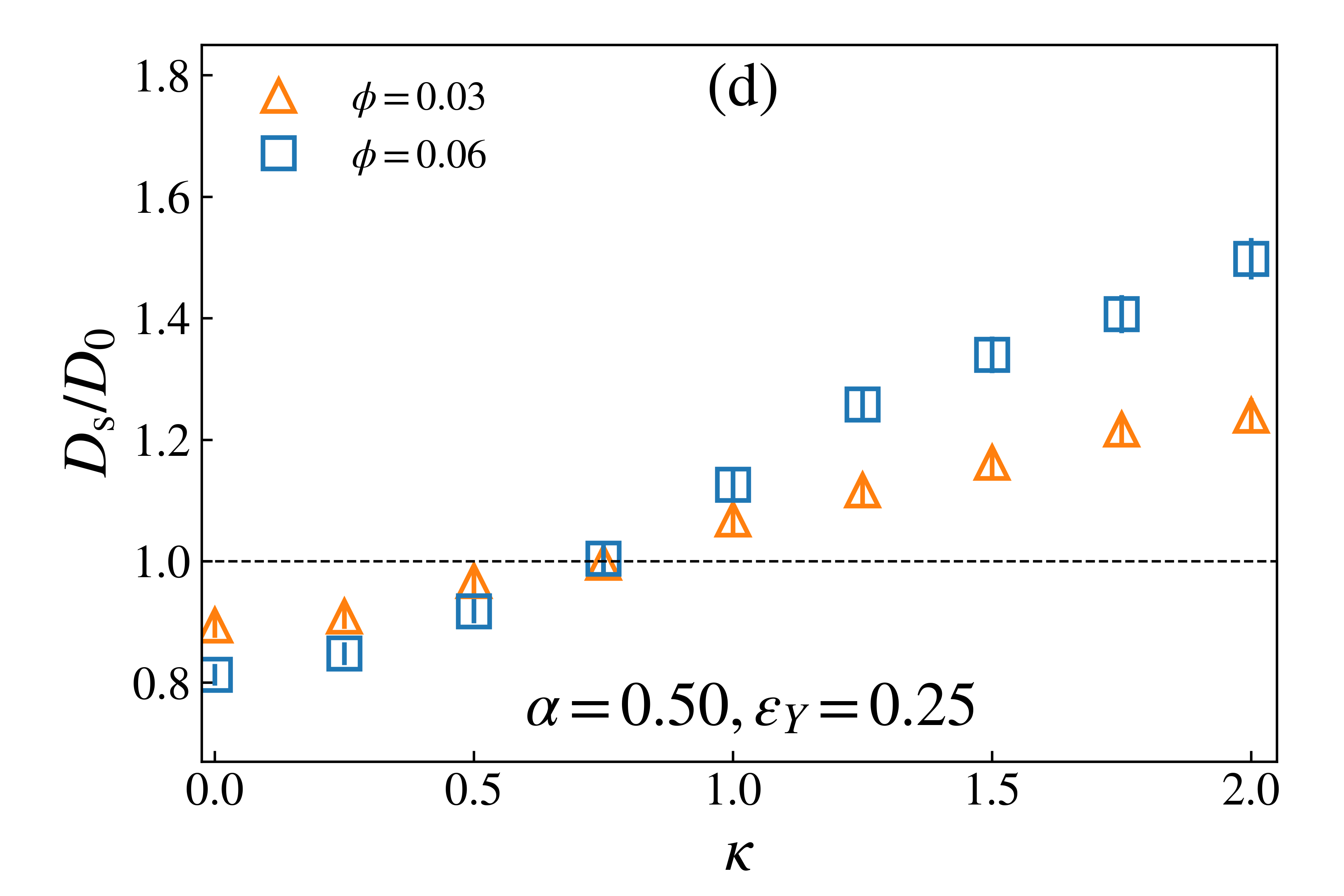}
\caption{(a,c): Reduced self-diffusion coefficient $D_\text{s}/D_0$ as a function of the area fraction $\phi$ for different values of the odd-diffusivity parameter $\kappa$. For $\kappa = 1$ the self-diffusion coefficient increases with increasing $\phi$ and for $\kappa = 0.5$ and $\kappa = 0.2$ it decreases with increasing $\phi$. This is to be compared with the reference case of normal particles, i.e. $\kappa = 0$. (b,d): $\kappa$-governed crossover from a reduced to an enhanced self-diffusion for two different area fractions. The inter-particle interaction is modeled with a Yukawa potential $U_\mathrm{Y}(r) = \varepsilon_Y/ r\ \exp\left[-(\sigma - r)/\alpha\right]$ additionally to the hard-core potential. (a,b) and (c,d) correspond to different parameters choices of the potential. All symbols represent data from Brownian dynamics simulations of hard-core particles diffusing under Lorentz force.}
\label{fig:yukawa}
\end{figure*}

\section{Simulation Details}
\label{section_simulations}

In the simulations we approximate the inter-particle hard-core repulsive force between particle
$i$ at position $\rr_i$ and particle $j$ at position $\rr_j$ by
\begin{align}
\f_{ij} = 
\left\{
\begin{tabular}{ll}
 $ \varepsilon \left( \frac{\sigma}{|\rr_{ij}|}\right)^\alpha \rr_{ij}$,~~
  & if ~~ $|\rr_{ij}| < \sigma_c$ \\
 0, & otherwise,
\end{tabular}
\right.
\end{align}
where
$\rr_{ij} = \rr_i - \rr_j$,
$\varepsilon$ sets the energy scale, $\sigma$ is the diameter of the particles,
$\alpha$ sets the steepness of the potential,
and $\sigma_c$ is the range of the potential.

In Figs.~\ref{fig:yukawa}(a)-(d), a Yukawa potential is included in addition to the hard-core interactions.
The force from the Yukawa potential is $\f^\mathrm{Y}_{ij} = -\nabla_i U_\mathrm{Y}(r_{ij})$, where
\begin{align}
U_\mathrm{Y}(r_{ij}) =
    \frac{\varepsilon_\mathrm{Y} }{r_{ij}} \exp[ - (\sigma- r_{ij}) / \alpha_\mathrm{Y}],
\end{align}
where $\varepsilon_\mathrm{Y}$ sets the energy at $r_{ij} = |\mathbf{r}_{ij}| = \sigma$ and $\alpha_\mathrm{Y}$ sets the decay length of the potential.
The total force on particle $i$ is 
\begin{align}
\F_i(\{\rr_i\}) =  \sum_{j \neq i}^N(\f_{ij} + \f^\mathrm{Y}_{ij}),
\end{align}
where $\f_{ij}$ is the force on the particle due to hard-core interaction and $N$ is the total number of particles in the system.

The stochastic differential equations used for the simulations are
\begin{align}
\frac{\partial \rr_i}{\partial t} &= \vv_i,
\label{r_sde}\\
\frac{\partial \vv_i}{\partial t}
&=
  - \frac{\gamma_i}{m_i} \left( \I - \kappa_i \boldsymbol{\epsilon} \right) \vv_i
  + \frac{1}{m_i} \F_i 
  + \frac{1}{m_i} \sqrt{2 \gamma_i T } \xxi_i,
\label{v_sde}
\end{align}
where
$T$ is the temperature in units such that the Boltzmann constant is unity, and $\gamma_i$, $m_i$ and $\kappa_i$ are, respectively,
the friction constant, mass and odd-diffusivity constant of particle $i$, and $\xxi_i$ is Gaussian white noise with $\left< \xxi_i\right> = 0$
and $\left< \xxi_i(t) \xxi_j(t')\right> = \I \delta_{ij} \delta(t - t')$.
Even though the friction matrix $\gamma_i ( \I - \kappa_i \boldsymbol{\epsilon})$
is not diagonal, the off-diagonal part proportional to $\boldsymbol{\epsilon}$ rotates the velocity without changing its
norm, and therefore does not dissipate energy.

The corresponding Fokker-Planck equation is
\begin{align}
\frac{\partial \mathcal{P}(t)}{\partial t} =&
-\sum_i^N 
\nabla_i \cdot \left[ \vv_i \mathcal{P}(t) \right] \nonumber\\
&+\sum_i^N \nabla_{\vv_i} \cdot \left[
\frac{\gamma_i}{m_i} \left( \I - \kappa_i \boldsymbol{\epsilon} \right) \vv_i
 \mathcal{P}(t) \right. \nonumber \\
&~~~~~~~~~~~~~~~~
-\frac{\F_i}{m_i} \mathcal{P}(t)
\left. + \frac{\gamma_i T}{m_i^2} \nabla_{\vv_i} \mathcal{P}(t)
\right],
\end{align}
where $\mathcal{P}(t)\equiv \mathcal{P}(\rr_1, ..., \rr_N, \vv_i, ... ,\vv_N, t)$.
In the overdamped limit ($\forall i\colon m_i/\gamma_i \rightarrow 0$),
and after integrating out the velocity degrees of freedom,
this becomes~\cite{chun2018emergence}
\begin{align}
\frac{\partial P(t)}{\partial t} = 
-\sum_i^N \nabla_i   \cdot \left[\mathbf{D}_i
\left( \frac{1}{T}\F_iP(t) -  \nabla_i P(t) \right) \right],
\end{align}
where
$P(t) \equiv P(\rr_1, ... ,\rr_N, t)$, and
$
\mathbf{D}_i = D_0^{(i)} \left( \I + \kappa_i \boldsymbol{\epsilon} \right)
$
is the diffusion tensor of particle $i$, with
$D_0^{(i)} = \frac{T}{\gamma_i} \frac{1}{1 + \kappa_i^2}$.
In the theory the forces are taken into account by using the appropriate 
boundary conditions (see Eq.~\eqref{bc_for_general_FP_eq})
and therefore do not appear explicitly in Eq.~\eqref{general_FP_eq} .

\subsection{Algorithm}
To numerically integrate Eqs.~\eqref{r_sde} and \eqref{v_sde}
we use an algorithm based on the forward Euler method.
Algorithm \ref{algorithm} shows the method to calculate
$\rr_i(t_{n+1})$ and $\vv_i(t_{n+1})$ from 
$\rr_i(t_{n})$ and $\vv_i(t_{n})$, with $t_{n+1} = t_n + \Dt$.
In step 1 the forces on the particles are calculated.
In step 3 the momentum change on particle $i$ due to
the friction, the forces and the thermal noise is calculated.
The thermal noise is accounted for by $\N(0,\sqrt{2 \gamma_i T \Dt})$,
which is a random vector drawn from a normal distribution with zero mean
and standard deviation $\sqrt{2 \gamma_i T \Dt}$.
In step 4 the momentum change due to the
$\gamma_i \kappa_i \boldsymbol{\epsilon} \vv_i$ part of Eq.~\eqref{v_sde}
is calculated.
This is done separately because this represents a rotation and therefore should
not change the length of the velocity vector.
The discrete nature of the algorithm causes a small change in the length after
the rotation (step 5).
This error is removed in step 6.
In step 8 the positions are updated.

\begin{figure}[H]
\begin{algorithm}[H]
  \caption{Forward Euler method}
  \label{algorithm}
   \begin{algorithmic}[1]
    \State calculate $\F_1, \ldots ,\F_N$
    \For{ $i =1 ,\ldots,  N$ }
        \State $\Dp_i \gets - \gamma_i  \vv_i(t_n) \Dt
                  + \F_i \Dt
                  + \N \left( 0, \sqrt{2 \gamma_i T \Dt} \right)  $
        \State $\Delta \pp_{i}^{(\kappa)} \gets
                \gamma_i \kappa_i \boldsymbol{\epsilon} \vv_i(t_n) \Dt$
        \State $\vv_i(t_{n+1})
                \gets \vv_i(t_n) + \Dp_i^{(\kappa)} / m_i$
        \State $\vv_i(t_{n+1})
                \gets \vv_i(t_{n+1}) | \vv_i(t_n)| / | \vv(t_{n+1})|$
        \State $\vv_i(t_{n+1}) \gets \vv_i(t_{n+1}) + \Dp_i/m_i$
        \State $\rr_i(t_{n+1}) \gets \rr_i(t_n) + \vv_i(t_n) \Dt$
    \EndFor
   \end{algorithmic}
\end{algorithm}
\end{figure}

\subsection{Parameter values}
The simulation results shown in the main text are obtained from a system with $N = 200$ particles.
We use square simulation box with periodic boundary conditions and sides of length $L$,
where $L$ is determined by the volume fraction $\phi$:
$L = \sqrt{ \frac{\pi N}{4 \phi}}$.
Note that in Figs.~3(b,c) there are two species $(N = N_1 + N_2)$. Species one, as representing the tagged particle has $N_1 = 1$ particles, and consequently $N_2 = 199$ for the host species.
The friction constant of particle $i$ is
 $\gamma_i = \gamma_0/(1+\kappa_i^2)$, with $\gamma_0 = 1$.
 The mass of particles is 
 \begin{align}
     m = \frac{m_0 }{1 + \max\{ \kappa_i \}^2},
 \end{align}
 such that $m < m_0  = 10^{-2}$
 and the velocity autocorrelation time ($m/\gamma_i$) is smaller than  $m_0/\gamma_0 = 10^{-2}$ for all particles.
 The other relevant time scale in the system is the time between collisions.
 The distance between particles is $\sim \phi^{-1/2}$.
 The typical time to diffuse over that distance is $\sim \left( 2 D_0 \phi \right)^{-1}$,
 which is always much larger than the velocity autocorrelation time.

The temperature of all particles was $T=1$ such that the bare diffusivity
of all particles is 
\begin{align}
D_0^{(i)} = \frac{T}{\gamma_i ( 1 + \kappa_i^2)} = \frac{T}{\gamma_0} = 1.
\end{align}

The parameters for the steep repulsive potential were
$\sigma=1$,
$\sigma_c = 1.01$,
$\varepsilon = 100$, and
$\alpha = 17$.

The time step for the numerical integration was $\Dt = 10^{-5}$.
The system was simulated for $t_{sim} = 10^4$,
and the mean-squared displacement $\left< \rr(t) \cdot \rr(t+\tau)\right>$ was calculated using the algorithm from \cite{frenkel2001understanding} for $0\leq \tau \leq 50$ averaged over the time origins.
The diffusion constant is obtained by fitting the mean squared displacement for $x$ and $y$  coordinates to
$a + 2 D \tau$ for $10 \leq \tau \leq 50$, where $D$ is the diffusion constant.
Note that the $D$ obtained this way is already averaged due to the average over the time origins.

\begin{figure}
\centering
\includegraphics[width=\columnwidth]{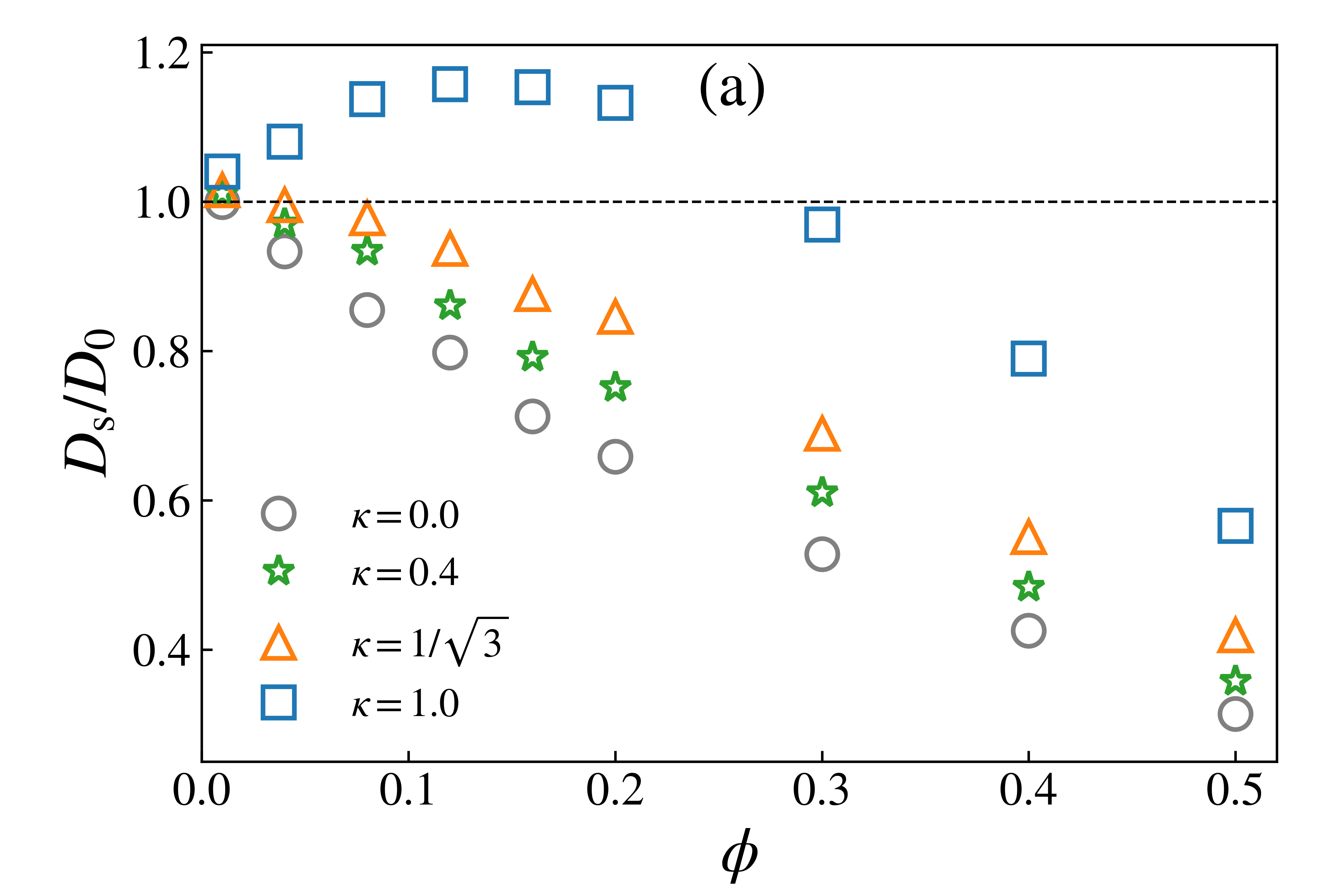}
\includegraphics[width=\columnwidth]{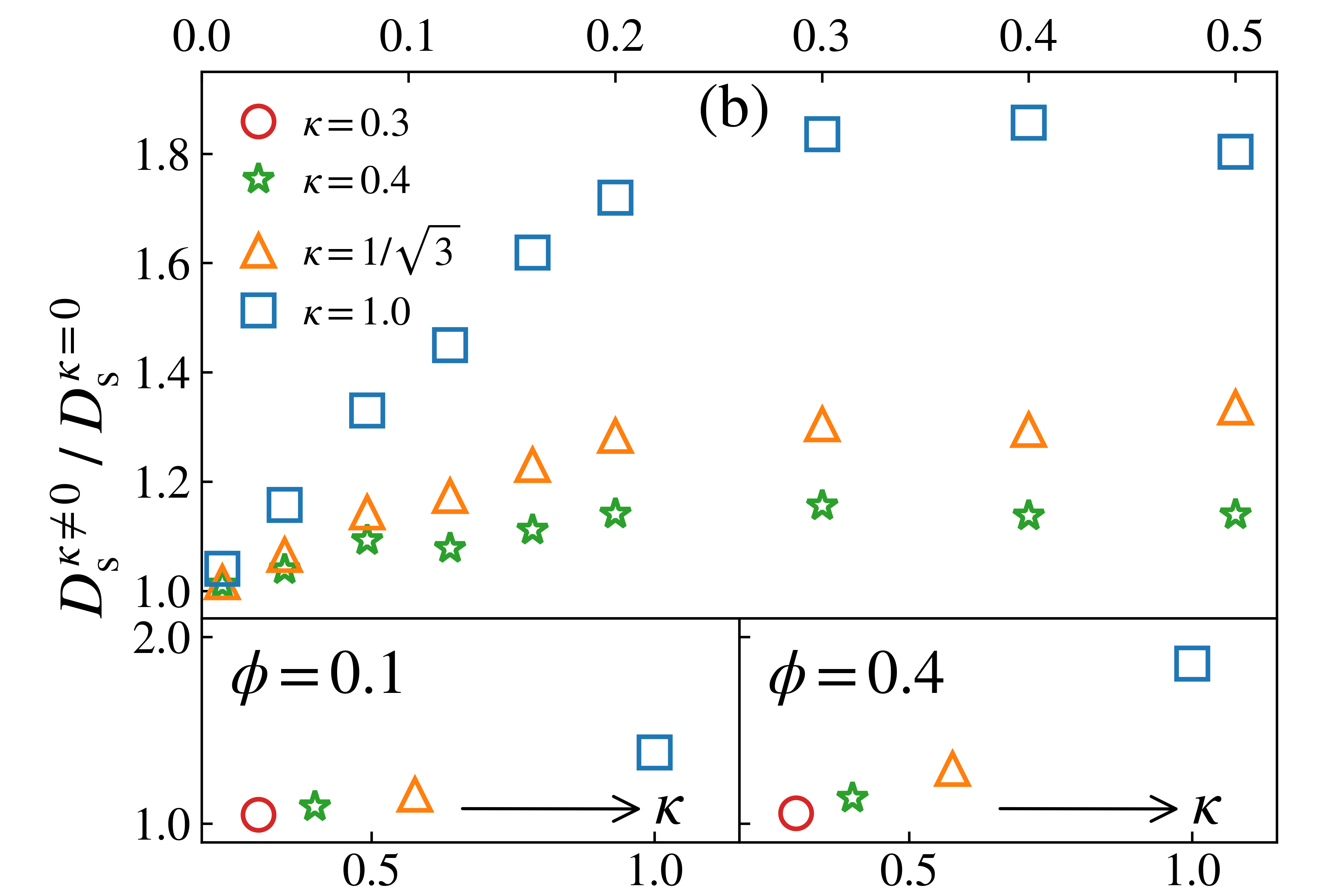}
\caption{(a) Reduced self-diffusion coefficient $D_\text{s}/D_0$ as a function of the area fraction $\phi$ for different values of the odd-diffusivity parameter $\kappa$. At large densities, the self-diffusion coefficient decreases due to many-body effects. The reduction in self-diffusion occurs for all values of $\kappa$. However, at all densities, the self-diffusion coefficient of an odd system ($\kappa \neq 0$) is larger than that of a normal system ($\kappa = 0$). (b) The relative enhancement of the self-diffusion due to odd-diffusivity is shown as a function of $\phi$. Apparently the relative enhancement increases with increasing $\kappa$. Odd-diffusivity also becomes more relevant with increasing density. The two figures below (b) show the $\kappa$ induced enhanced self-diffusion for two densities. All symbols represent data from Brownian dynamics simulations of hard-core particles diffusing under Lorentz force. Error bars are smaller than the symbols.}
\label{fig:high_density}
\end{figure}

For Figs.~2(a,b), and Fig.~3(a) of the main text and Figs.~\ref{fig:high_density}(a,b) this results in 400 samples of $D$ of the tagged particle
(200 particles, 2 coordinates per particle).
For Figs.~3(b,c) of the main text this results in only 2 samples (1 tagged particle, 2 coordinates per particle).
To increase the sample size for the latter case, the simulation was repeated 5 times with different random numbers, resulting in a total of 10 samples of $D$.
The average of the samples is reported in the figures, and the error bars represent estimate of the standard error
calculated as the standard deviation of the samples divided by the square root of the number of samples.

The data with the Yukawa potential shown in Figs.~\ref{fig:yukawa}(a)-(d), were done with the same parameters as in the data in Fig.~2 of the main text but with a total simulation time of $t_{sim} = 500$.

\section{Soft Interaction Potential}
\label{section_soft_interactions}

The advantage of modeling interactions via a hard-core potential is that they are amenable to analytical treatment. However, our main findings are not sensitive to the details of interactions. To demonstrate this, we performed Brownian dynamics simulations of odd-diffusive particles with soft interaction potentials. We specifically considered a Yukawa interaction potential $U_\mathrm{Y}(r) = \varepsilon_\mathrm{Y}/ r\ \exp\left[-(\sigma - r)/\alpha\right]$, where the parameter $\alpha$ is the decay length of the potential and $\varepsilon_\mathrm{Y}$ is the energy at $r = \sigma$. The results are shown in Fig.~\ref{fig:yukawa}. As in the case of hard-core interactions, collisions enhance the self-diffusion coefficient in a system with soft interactions. There exists a critical $\kappa$ above which, the self-diffusion coefficient increases with increasing area fraction of the host particles (Figs.~\ref{fig:yukawa}(a,c)). There is also a $\kappa$-governed crossover from a reduced to an enhanced self-diffusion at a fixed area fraction (Figs.~\ref{fig:yukawa}(b,d)). These results are qualitatively similar to a system with hard-core interactions.

\section{High Densities}
\label{section_high_densities}
 
In the main text, we studied the effect of interactions on the self-diffusion coefficient of an odd-diffusive particle in the low density limit. It is expected that the self-diffusion coefficient decreases at higher densities of host particles. Indeed we observe an apparent plateau in the self-diffusion which persisted up to $\phi \approx 0.2$. From this we conclude that even at larger densities ($\phi > 0.1$), the mutual rolling effect partially compensates for the slowing down due to the many-body effects. In this section, we show results for the self-diffusion coefficient up to $\phi = 0.5$ area fraction of (odd-diffusive) host particles. As shown in Fig.~\ref{fig:high_density}(a), the self-diffusion coefficient decreases with increasing $\phi$. However, the self-diffusion coefficient of an odd-diffusive system is always larger than that of a normal diffusing system ($\kappa = 0$). In other words, at all densities, an odd-diffusive system exhibits faster self-diffusion than a normal diffusing system. By considering the ratio of the self-diffusion coefficient of an odd-diffusive system to that of a normally diffusing system, we can quantify the relative enhancement of the self-diffusion coefficient due to the odd-diffusivity of the particles. This is shown in Fig.~\ref{fig:high_density}(b), where it is evident that the relative enhancement becomes stronger with increasing odd-diffusivity to the extent that at $\phi \approx 0.4$, the ratio exceeds $1.8$.

\bibliographystyle{unsrt}